\documentclass[%
 reprint,
 amsmath,amssymb,
 aps,
prb,
]{revtex4-1}

\usepackage{lettrine}
\usepackage{graphicx}
\usepackage{dcolumn}
\usepackage{bm}
\usepackage[maxfloats=256]{morefloats}
\usepackage{xcolor}
\begin{document}

\preprint{APS/123-QED}

\title{A semi-analytical approach to calculating the dynamic modes of magnetic vortices with Dzyaloshinskii-Moriya interactions}

\author{Carla Quispe Flores$^1$}
\author{Casey Chalifour$^2$}%
\author{Jonathon Davidson$^2$}%
\author{Karen L. Livesey$^2$}%
\author{Kristen S. Buchanan$^1$}%
\thanks{Kristen.Buchanan@colostate.edu}%
\affiliation{%
$^1$Department of Physics, Colorado State University, Fort Collins, USA}%
\affiliation{%
$^2$Department of Physics, University of Colorado Colorado Springs, Colorado Springs, USA
}%

\date{\today}

\begin{abstract}

Here we introduce a Landau-Lifshitz based diagonalization (LLD) method, and use this approach to calculate the effects of the interfacial Dzyaloshinskii Moriya interactions (DMI) on the radial-type spin wave modes of magnetic vortices in circular disks. The LLD method is a semi-analytical approach that involves the diagonalization of the magnetostatic  kernel, exchange, and DMI contributions to extract the system eigenfrequencies and eigenmodes.  The magnetic vortex state provides a convenient model system in which to investigate the effects of the DMI on the dynamics of a magnetic structures with confined geometries. Our calculations show that the DMI leads to shifts of the mode frequencies that are similar in magnitude to what is observed for spin waves of a comparable wavelength in extended films. However, unlike what is found in thin films, only the down-shifted modes are observed in the disks, and these corresponds to modes that propagate either radially outward or inward, depending on the vortex circulation. The semi-analytical calculations agree well with full micromagnetic simulations. This technique also applies to other systems with cylindrical symmetry, for example, magnetic skyrmions. 

\end{abstract}

\maketitle

\section{Introduction}
Interfacial Dzyaloshinskii-Moriya interactions (DMI) are important for the stabilization of N\'eel skyrmions \cite{Fert2013SkyrmionsTrack,Rohart2013SkyrmionInteraction,Woo2016ObservationFerromagnets,Moreau-Luchaire2016AdditiveTemperature}, and have also been shown to influence domain wall formation and propagation\cite{Ryu2013ChiralWalls,Emori2013Current-drivenWalls,DeJong2015AnalyticAnisotropy}. The DMI also lead to significant changes in the dispersion relations for surface spin waves in saturated magnetic thin films \cite{Moon2013Spin-waveInteraction,Kostylev2014InterfaceInteraction}, where inclusion of a DMI energy term results in shifts in the frequencies of surface spin waves that propagate in opposite directions. These theoretically predicted frequency shifts have been verified experimentally by several groups, and the detection of this frequency difference is currently the best available method to obtain quantitative measurements of the DMI \cite{Stashkevich2015ExperimentalInteraction,Nembach2015LinearFilms,Di2015DirectFilm,Ma2017Dzyaloshinskii-MoriyaInterface}. 
 
The DMI have intriguing potential for spintronics and magnonics applications, consequently it is important to develop a full understanding of how the DMI affect dynamics in patterned magnetic structures. The dynamic excitations in micro- and nano-sized structures are quantized based on the element size, hence the wavelengths of the spin excitations are short enough that they are likely to be affected by the DMI. In the presence of DMI, counterpropagating surface spin waves that have the same frequency will have different wavelengths and, as a direct consequence, the spin excitations in confined geometries can no longer be standing wave excitations \cite{Zingsem2019UnusualWaves}. Micromagnetic simulations have also shown that the DMI should lead to non-reciprocal modes that propagate around the edges of saturated structures \cite{Garcia-Sanchez2014NonreciprocalInteraction}, and to changes of the spin eigenmodes of nonuniform magnetization states \cite{Mruczkiewicz2017SpinStates}, where the dynamic spectra for spin textures ranging from a magnetic skyrmion to a vortex were considered as the anisotropy and DMI were varied. Magnetic vortices, in particular, have often served as a model system for the study of spin textures. In the dynamic regime, vortices exhibit modes with gyrotropic \cite{Novosad2005MagneticDots}, radial \cite{Demokritov2008Micro-BrillouinNanostructures,Vogt2011OpticalDisks}, and azimuthal \cite{Giovannini2004SpinStates} motion patterns. The vortex radial modes, which have been described theoretically \cite{Guslienko2005Vortex-stateDots,Zaspel2009FrequenciesDisks} but without the inclusion of the DMI, involve spin waves that are quantized in the radial direction, also the direction that should be maximally affected by interfacial DMI. Hence this is a convenient system in which to examine the effects of the DMI on spin textures. 

The paper is structured as follows: In section II, we introduce the Landau-Lifshitz based diagonalization (LLD) method to solve for the spin excitation frequencies and mode profiles for spin states with cylindrical symmetry and with DMI. In section III, the approach used for micromagnetic simulations is presented. In  sections IV, we show that the LLD method and micromagnetic simulations yield frequencies and mode profiles that agree well for a magnetic vortex state. The results show that the DMI lead to important modifications of the spin excitations in patterned structures. The modifications of the vortex dynamics are illustrative of what should be expected for other spin configurations, and this approach can be easily extended to other systems with cylindrical symmetry, e.g., skyrmions. Finally, in section V we provide the conclusions.

\section{Theory}
We use a semi-analytical approach to study the effects of the DMI on the eigenfrequencies and eigenmodes of magnetic vortices confined in cylindrical nanodots of thickness $L$ and radius $R$. Radial-type modes with frequencies well above the vortex core gyrotropic frequency are considered (typically radial modes are in the GHz range, whereas the gyrotropic mode $\sim100$~MHz), and we assume that the magnetization does not depend on the $z$-coordinate through the disk thickness. Further, the DMI is of the interfacial type with symmetry breaking along the $z$ direction as observed in heavy metal/ferromagnetic bilayer or multilayer thin films. 

\begin{figure}[h!]
\includegraphics[width=\columnwidth]{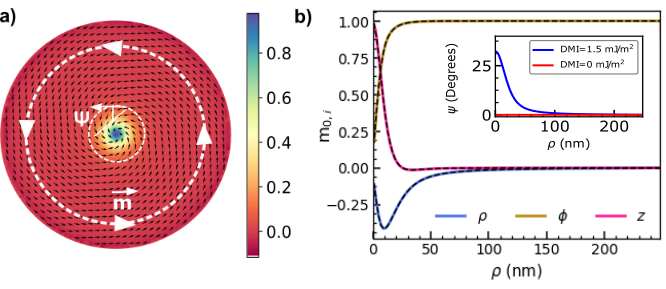}
\caption{(Color online) Static vortex configuration of a disk of radius $R = 250$~nm with $D=1.5$~mJ/m$^2$, where a) shows a top view of the spin distribution where the colorbar represents $m_{o,z}$, and b) depicts the magnetization components vs.~$\rho$, where solid lines are from micromagnetic simulations, and dashed lines are solutions obtained from 1D energy minimization. The spin state has cylindrical symmetry, but the spins are tilted by an angle $\psi$ away from the $\hat{\phi}$ direction, shown in the inset. For $D=0$~mJ/m$^2$, $\psi$ is zero.}
\label{fig:vortex_tilt}
\end{figure} 
The magnetic energy for a cylindrical disk including the exchange, magnetostatic and DMI energy terms is
\begin{multline}
\label{W}
W = -2\pi L\int^R_0 \rho\,d\rho\, [ A_{ex} \mathbf{m} \cdot \nabla^2 \mathbf{m} \\  -\frac{\mu_o}{2}M_s^2(\mathbf{h_d}+\mathbf{h_{DM}})\cdot \mathbf{m} ],
\end{multline}
\noindent
where $\rho$ is the radial coordinate, $\mathbf{m}=\mathbf{M}/M_s$ is the reduced magnetization, $M_s$ is the saturation magnetization, $A_{ex}$ is the exchange stiffness constant, and $\mathbf{h_d}$ and $\mathbf{h_{DM}}$ are the reduced effective demagnetization and DMI fields, respectively, normalized by $M_s$. Anisotropy is neglected, since vortices form most readily in magnetically soft materials where the anisotropy is small in comparison with the magnetostatic energy of the vortex. We note that interfacial DMI usually co-exists with some level of out-of-plane anisotropy. A vortex state will exist for low out-of-plane anisotropy $K_u$, whereas for larger $K_u$, perpendicular domains or skyrmions will form depending on the DMI. Provided the anisotropy is small enough to maintain a vortex state, the anisotropy will not appreciably change the main characteristics of the excitations and the approximate effect of $K_u$ is a reduction of the effective magnetization extracted from a dynamic measurement by $2K_u/\mu_o M_s$. 

\begin{figure}[t]
    
	\includegraphics[width=\columnwidth]{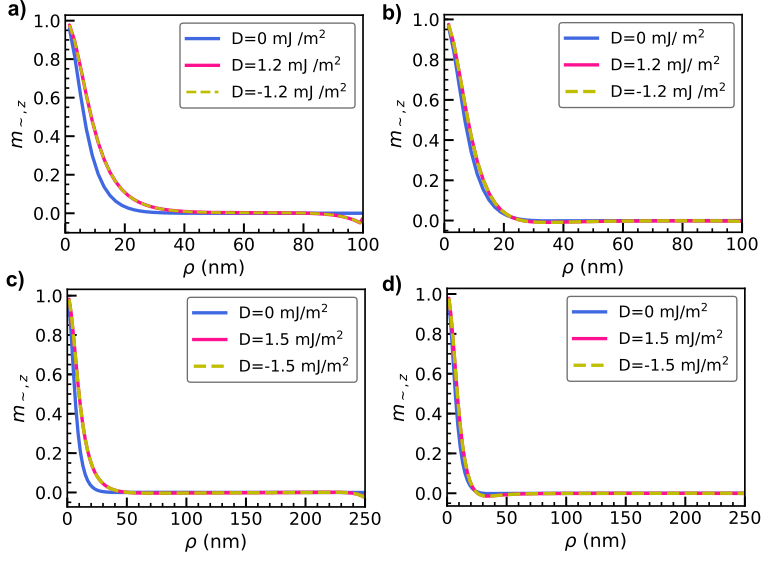}
    \caption{Static $m_{0,z}$ profiles for three representative $D$ values show an enlargement of the vortex core and an increasingly prominent out-of-plane tilt of the edge magnetization with increasing $D$. The disk dimensions are a) $R=100$~nm and $L=1$~nm , b) $R=100$~nm and $L=5$~nm, c)$R=250$~nm and $L=1$~nm, and d) $R=250$~nm and $L=5$~nm.
    }
    \label{fig:COREmz}
\end{figure}

The DMI between two atomic spins $\mathbf{S}_i$
and $\mathbf{S}_j$ is given by 
\begin{align}
\label{Atomistic-dm}
\mathcal{H}_{DMI} = \mathbf{D}_{ij} \mathbf{\cdot}(\mathbf{S}_i \times \mathbf{S}_j),
\end{align}
where $\mathbf{D}_{ij}$ is the Dzyaloshinskii-Moriya vector, which is perpendicular to both the asymmetry direction and the vector $\mathbf{r}_{ij}$ between the spins $\mathbf{S}_i$ and $\mathbf{S}_j$. This atomistic model translated to a continuum energy density in cylindrical coordinates with symmetry breaking along the $\mathbf{ \hat{z}}$ direction, reads

\begin{multline}
\label{dm_energy_density}
E_{DM} = -D \left[ (\hat{\phi} \times \hat{z}) \cdot \left( \mathbf{m} \times \frac{1}{\rho} \frac{ \partial \mathbf{m}}{\partial \phi} \right)  \right. + \\  
\left.
( \hat{\rho} \times \hat{z})\cdot \left(\mathbf{m} \times \frac{\mathbf{m}}{\partial \rho}\right) \right].
\end{multline}
If the magnetization is only a function of $\rho$, the DM energy density reduces to
\begin{align}
\label{dm_vtx_energy_density}
E_{DM} =  -D \left[ m_{\rho} \frac{\partial m_z}{\partial \rho}  -m_z \left( \frac{\partial m_{\rho}}{\partial \rho}  + \frac{m_{\rho}}{\rho}\right)\right],
\end{align}
with the associated effective field 
\begin{align}
\label{dm_heff}
\mathbf{h}_{DM} =  \frac{2D}{\mu_o M_s^2} \left[ \frac{\partial m_z}{\partial \rho} \hat{\rho} - \left( \frac{\partial m_{\rho}}{\partial \rho}  + \frac{m_{\rho}}{\rho}\right)\hat{z}\right].
\end{align}
\color{black}
Since the vortex magnetization has radial symmetry, an assumption that is supported by micromagnetic simulations for $|D|\geq0$~mJ/m$^2$, the two-dimensional spin distribution is effectively reduced to a one-dimensional system parameterized by the radial coordinate $\rho$. Under this simplification, we set the magnetization to a random state and find the static configuration $\mathbf{m}_0(\rho)$ by minimizing the disk energy, Eq.~(\ref{W}), using the nonlinear conjugate gradient method with the modified Polak-Ribiere-Polyak-method with guarantee conjugacy \cite{Fischbacher2017NonlinearMicromagnetics}. The results, shown in Fig.~\ref{fig:vortex_tilt}, indicate that the static magnetization of the vortex-state with DMI is tilted from the azimuthal direction by an angle $\psi=\tan^{-1}{\left(-m_\rho/m_\phi \right)}$  that is well described by a Cauchy function for most values of DMI and dot aspect ratios, $\psi(\rho) = \frac{a}{1+ \left(\frac{\rho}{\sigma R}\right)^2}$,  where $a$ and $\sigma$ are fitting parameters. The tilt $\psi$ goes to zero as $D$ approaches zero, and it also disappears for a ring, which occurs because a vortex in a ring lacks an out-of-plane core and there is consequently no DMI energy advantage to a tilt. A small out-of-plane canting of the spins also develops at the disk edge due to the DMI, as Fig.~\ref{fig:COREmz} shows.

To describe small oscillations of the magnetization about $\mathbf{m}_0(\rho)$, we write the Landau-Lifshitz (LL) equation in a new orthogonal system $(s, \xi,z)$, where the magnetization is rotated around the $z$ axis at each $\rho$ position by the tilt angle $\psi{(\rho)}$ using the rotation matrix

\begin{align}
\label{R}
\mathcal{R}(\psi)= \begin{pmatrix}
\cos\psi & -\sin \psi & 0 \\ 
\sin \psi & \cos \psi & 0 \\ 
 0 & 0  & 1
\end{pmatrix}
\end{align}
such that the reduced magnetization and reduced total effective field in the rotated frame are
\begin{align}
\mathbf{\widetilde{m}}= \mathcal{R}(\psi) \mathbf{m},\\
\mathbf{\widetilde{h}} = \mathcal{R}(\psi) \mathbf{h},
\end{align}
respectively, with

\begin{align}
\mathbf{m}=
\begin{pmatrix}
m_{\rho} \\ 
m_\phi\\
m_z
\end{pmatrix}, 
\mathbf{\widetilde{m}}=
\begin{pmatrix}
\widetilde{m}_s \\ 
\widetilde{m}_{\xi}\\
m_z
\end{pmatrix}, 
\mathbf{h}=
\begin{pmatrix}
h_{\rho} \\ 
h_\phi\\
h_z
\end{pmatrix}, \textrm{and}~
\mathbf{\widetilde{h}}=
\begin{pmatrix}
\widetilde{h}_s \\ 
\widetilde{h}_{\xi}\\
h_z
\end{pmatrix}.
\end{align}

The LL equation in the rotated frame with no damping reads
\begin{align}
\label{LLrot}
 \frac{ \partial \widetilde{\mathbf{m}}}{\partial t}=-|\gamma| M_s \mathbf{ \widetilde{m} \times \widetilde{h}},
\end{align}
where $\gamma$ is the gyromagnetic ratio, and the reduced total effective field in the rotated frame $\mathbf{\widetilde{h}}$,  
\begin{align}
\mathbf{ \widetilde{h}} = \mathcal{R}(\psi)(\mathbf{h_{ex}}+ \mathbf{h_d}+\mathbf{h_{DM}}),
\end{align}
includes the exchange $\mathbf{h_{ex}}$, demagnetization $\mathbf{h_{d}}$, and DMI $\mathbf{h_{DM}}$ fields. The magnetization and the effective field can each be separated into static and dynamic parts
\begin{equation*}
\begin{aligned}
\mathbf{ \widetilde{m}} = \mathbf{\widetilde{m}_0}(\rho) + \bm{\widetilde{m}_{\sim}}(\rho,t), \\
\mathbf{\widetilde{h}} = \mathbf{\widetilde{h}_0} (\rho) + \mathbf{\widetilde{h}_{\sim}} (\rho,t), 
\end{aligned}
\end{equation*}

\noindent
where the dynamic magnetization vector 
$\bm{\widetilde{m}_{\sim}}(\rho,t)=(\widetilde{m}_{\sim,s},0,m_{\sim,z})$ 
is perpendicular to the static magnetization vector $\mathbf{\widetilde{m}_0}=(0,1,0)$ that is assumed to lie in-plane. This is a reasonable assumption since the out-of-plane vortex core is small (typically $\sim$10~nm), and calculations that include the out-of-plane tilt yield similar results. The LL equation is linearized considering
that $|\mathbf{m}_{\sim}|<<|\mathbf{m}_o|$,~$\mathbf{\widetilde{m}_0}\times \mathbf{\widetilde{h}_{0}}=0$, and the temporal variation of the dynamic components is assumed to be of the form $\exp(-i \omega t)$, leading to 
\begin{align}
\label{LL-Lineal}
-i \omega \bm{\widetilde{m}_\sim} =  -|\gamma| M_s ( \mathbf{\widetilde{m}_0}\times \mathbf{ \widetilde{h}_\sim} + \bm{\widetilde{m}_\sim}\times \mathbf{\widetilde{h}_0}),
\end{align}
which yields a set of two coupled equations
\begin{align}
i\omega \widetilde{m}_{\sim,s} = |\gamma|M_s (\widetilde{m}_0 h_{\sim,z} -m_{\sim,z} \widetilde{h}_{0,\xi}) \\
i\omega \widetilde{m}_{\sim,z} = |\gamma|M_s (-\widetilde{m}_0 \widetilde{h}_{\sim,s} + \widetilde{m}_{\sim,s} \widetilde{h}_{0,\xi}),
\end{align}
where terms that involve products of dynamic contributions are neglected. 

The demagnetizing field contribution is $\mathbf{h}_d(\mathbf{r}) = \widehat{G} [\mathbf{m}(\mathbf{r})]$ where $\widehat{G}$ is a tensorial non-local integral operator (the tensorial magnetostatic Green's function $G_{\alpha\beta}(\mathbf{r,r^{\prime}}) = -(\nabla_{\mathbf{r}})_{\alpha} (\nabla_{\mathbf{r^{\prime}}})_{\beta} (4\pi |\mathbf{r}-\mathbf{r^{\prime}}|)^{-1} $) expressed in cylindrical coordinates with $\mathbf{r}=(\rho,\phi,z)$ as $\widehat{G} [\mathbf{m}(\mathbf{r})] = \int \widehat{G}[\mathbf{r,r^{\prime}}] \mathbf{m(\mathbf{r^{\prime}})} d^3 \mathbf{r^{\prime}}$. Due to the radial symmetry and constant magnetization $\mathbf{m(r)}$ across the disk thickness, the Green's functions can be averaged over $\phi\phi^{\prime}$ and $zz^{\prime}$
\begin{align}
g_{\alpha\beta}(\rho,\rho') = \frac{1}{2 \pi L} \int^L_0 dz \int^{2\pi}_{0}d\phi \int^L_0 dz^{\prime} \int^{2\pi}_{0}d\phi^{\prime}G_{\alpha \beta}(\mathbf{r,r^{\prime}}).
\end{align}
The only terms that are nonzero are
\begin{widetext}
\begin{align}
\label{grdemag}
g_{\rho \rho} (\rho, \rho')= - \frac{1}{\rho'}\delta(\rho -\rho')+\frac{(2-\gamma ^{2})K_{ell}\left [\gamma ^{2}  \right ]-2E_{ell}\left [\gamma ^{2}  \right ]}{L\pi\gamma \sqrt{\rho \rho'}} -\frac{(2-\gamma _{L}^{2} )K_{ell}\left [\gamma _{L}^{2}  \right ]-2E_{ell}\left [\gamma _{L}^{2} \right ]}{L\pi\gamma_{L} \sqrt{\rho \rho'}} 
\end{align}
\begin{align}
\label{gzdemag}
g_{zz} (\rho, \rho') =-\frac{2}{ \pi L} \left \{ \frac{1}{\rho'}K_{ell}\left [ \frac{\rho ^{2}}{{\rho'} ^{2}} \right ]\left ( 1-\Theta \left [ \rho -\rho' \right ] \right ) +\frac{1}{\rho }K_{ell}\left [ \frac{{\rho'} ^{2}}{\rho ^{2}} \right ]\Theta \left [ \rho -\rho' \right ]-\frac{1}{\gamma _{2}}K_{ell}\left [ \frac{\gamma _{1}^{2}}{\gamma _{2}^{2}} \right ]\right \}
\end{align}
\end{widetext}
where $\Theta[\rho -\rho']$ is the Heaviside step function, $K_{ell},E_{ell}$ are elliptic integrals, and
\begin{align}
\gamma= \sqrt{ \frac{4 \rho \rho^{\prime}}{(\rho+\rho')^2}}, \ \ \gamma_L= \sqrt{ \frac{4 \rho \rho^{\prime}}{L^2+(\rho+\rho')^2}}, \\
\gamma_{1,2}=\frac{1}{2} \left[  \sqrt{(\rho+\rho')^2+L^2} \mp \sqrt{(\rho -\rho')^2+L^2} \right].
\end{align}
It is important to mention that Eqs.~(\ref{grdemag}) and~(\ref{gzdemag}) can also be used for the dynamic calculations since we are in the magnetostatic regime, i.e., $\omega<<ck$, where $k$ is the wave vector.

The demagnetization field in cylindrical coordinates can now be reduced to 
\begin{align}
\label{hd-3}
\mathbf{h_d}(\rho) = \hat{\Gamma}_d [\mathbf{m(\mathbf{\rho^{\prime}})}]
\end{align}
where $\hat{\Gamma}_d$ is the magnetostatic tensorial non-local integral operator, which operates on $ \mathbf{m}(\mathbf{\rho^{\prime}})$ as follows
\begin{align}
\label{hd-2}
\hat{\Gamma}_d [\mathbf{m(\mathbf{\rho^{\prime}})}] = \int  \hat{g}(\rho,\rho^{\prime}) \mathbf{m(\mathbf{\rho^{\prime}})} \rho' d\rho',
\end{align}
with
\begin{align}
\hat{\Gamma}_d  = \begin{bmatrix}
\hat{A}_{\rho \rho} & 0&   0 \\ 
0 & 0&   0 \\
0  & 0&\hat{A}_{z z},
\end{bmatrix}
\end{align}
\begin{align}
\label{Arr}
\hat{A}_{\rho \rho} =  \hat{g}_{\rho \rho} \ \ \mathrm{diag}( \rho') \ \  \Delta\rho', \\
\label{Azz}
\hat{A}_{zz} =  \hat{g}_{zz} \ \ \mathrm{diag}( \rho') \ \  \Delta\rho',
\end{align}
where the non-local integral operators $\hat{g}_{\rho\rho}$ and $\hat{g}_{zz}$ are written in $n \times n$ discretized matrix form with $\rho$ entries as columns, and $\rho'$ entries as rows following Eqs.~(\ref{grdemag}-\ref{gzdemag}), $\mathrm{diag}(\rho')$ is an $n \times n$ diagonal matrix, and $\Delta\rho'$ is the cell size. Note that the integration in Eq.(\ref{hd-2}) is accomplished by the matrix multiplication between the non-local integral operator $\hat{\Gamma}_d$ with the column vector $\mathbf{m(\mathbf{\rho^{\prime}})}$ (for more details see the Appendix: Definitions of matrix operators). 

The DMI Eq. (\ref{dm_heff}) and exchange effective fields can be expressed in terms of differential operators as
\begin{align}
\mathbf{h_{DM}}(\rho) = \hat{\Gamma}_{DM}\mathbf{m(\mathbf{\rho})}, \ \
\mathbf{h_{ex}}(\rho) =\hat{\Gamma}_{ex} \mathbf{m(\mathbf{\rho})}, 
\end{align}
with 
\begin{align}
\hat{\Gamma}_{DM}= \begin{pmatrix}
\hat{D}_{\rho \rho}^{BC} & 0 & \hat{D}_{\rho z}\\ 
0 & 0 & 0\\
\hat{D}_{z \rho} & 0 &\hat{D}_{z z}^{BC}
\end{pmatrix}, \hat{\Gamma}_{ex}=\begin{pmatrix}
\hat{E}_{\rho \rho} & 0 & \hat{E}_{\rho z}^{BC} \\ 
0 & \hat{E}_{\phi \phi} &0 \\
 \hat{E}_{z \rho}^{BC} & 0&\hat{E}_{z z}
\end{pmatrix}.
\end{align}
The operators with the superscript $BC$ contain only terms that contribute to the  boundary conditions and in the absence of DMI these terms vanish. For spin textures with radial symmetry the extended form of the above operators are 
\begin{gather}
\hat{D}_{\rho z} = \frac{2 D}{\mu_0 M_s^2} \frac{d }{d \rho} \\
\hat{D}_{z \rho} = -\frac{2 D}{\mu_0 M_s^2} \left(\frac{d}{d \rho} +\frac{1}{\rho}\right)\\
\hat{E}_{\rho \rho}= \hat{E}_{\phi \phi}= \frac{2 A_{ex}}{\mu_0M_s^2}\left( \frac{d^2}{d \rho^2} +\frac{1}{\rho} \frac{d}{d \rho} -\frac{1}{\rho^2}\right) \\
\hat{E}_{zz}= \frac{2 A_{ex}}{\mu_0M_s^2}\left( \frac{d^2}{d \rho^2} +\frac{1}{\rho} \frac{d}{d \rho} \right) 
\end{gather}
subject to the boundary conditions
\begin{align}
\label{dmi-BC}
\frac{d\mathbf{m}}{dn} = \frac{D}{2A_{ex}}(\hat{z} \times \hat{n})\times \mathbf{m}.
\end{align}
This form guarantees that the edge magnetization rotates in a plane containing the edge surface normal \cite{Rohart2013SkyrmionInteraction}.

\begin{figure}[b]
\centering
\includegraphics[width=\columnwidth]{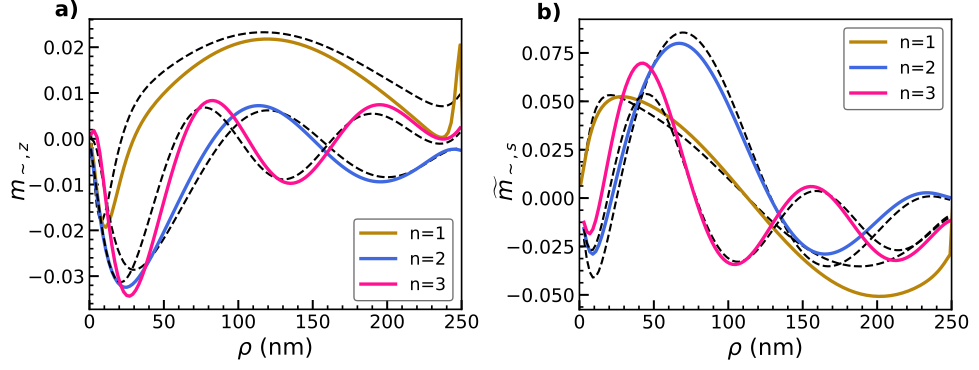}
\caption{(Color online) Eigenvectors a) $m_{\sim,z}$ and b) $\widetilde{m}_{\sim,\rho}$ corresponding to the first three vortex radial modes for $R=250$~nm, $L=5$~nm, and $D=1.5$~mJ/m$^2$. The solid lines are from the simulations, while the dashed lines show the LLD semi-analytical solutions. The eigenfrequencies obtained from the LLD approach are $f_1=(5.15+0.0019i)$~GHz, $f_2=(6.51+0.0012i)$~GHz, and $f_3=(7.73+0.003i)$~GHz.}
\label{fig:mzmrEigenvector}
\end{figure}

 Semi-analytical solutions were obtained by extending the approach presented in Ref.~[\onlinecite{Guslienko2005Vortex-stateDots}] for the case without DMI. In the absence of DMI, the linearized LL equation Eq.(\ref{LL-Lineal}) can be reduced to a single integro-differential equation and then solved numerically as an eigenvalue problem. The effective torque exerted by the DMI, however, yields two 
 uncoupled integro-differential equations, hence the problem is consequently more complicated than the dipolar-only case. With DMI, the problem can be set up as an eigenvalue problem of the form
\begin{align}
\label{Eigenvealu-problem}
\frac{i \omega}{|\gamma| M_s} \bm{m}_{\sim,n} = \hat{\Gamma}(\rho,\rho^{\prime}) \bm{m}_{\sim,n}
\end{align}
with eigenfrequencies obtained as follows
\begin{align}
\omega_n = -i \gamma M_s \lambda_n,
\end{align}
where $\bf{m}_{\sim,n}(\rho)$ is the $n^{th}$ eigenmode with eigenfrequency $\omega_n$, and $\lambda_n$ are the eigenvalues of the operator $\hat{\Gamma}$, which contains the magnetostatic nonlocal integral operator, and the exchange and DMI differential operators.
The extended form of Eq.~(\ref{Eigenvealu-problem}) reads
\begin{widetext}
\begin{align}
\label{Eigenvealu-problem-extended}
\frac{i \omega}{ |\gamma| M_s} 
\begin{pmatrix}
\widetilde{m}_{\rho}\\ 
m_z
\end{pmatrix} = 
\begin{pmatrix}
\widetilde{m}_{0} \hat{\Gamma}_{z\rho}\cos \psi &  \widetilde{m}_{0} \hat{\Gamma}_{zz} -\tilde{h}_{o,\xi}\\ 
\widetilde{h}_{o,\xi} - \widetilde{m}_{0} \left( \hat{\Gamma}_{\rho \rho} \cos^2 \psi + \hat{\Gamma}_{\phi \phi} \sin^2 \psi \right)&  -\widetilde{m}_{0}\hat{\Gamma}_{\rho z}\cos\psi 
\end{pmatrix}
\begin{pmatrix}
\widetilde{m}_{\rho}\\ 
m_z
\end{pmatrix},
\end{align}
\end{widetext}
with 
\begin{align}
\hat{\Gamma}_{\rho \rho}= \hat{A}_{\rho\rho} + \hat{E}_{\rho\rho}+\hat{D}^{BC}_{\rho\rho} \\
\hat{\Gamma}_{zz}= \hat{A}_{zz} + \hat{E}_{zz}+\hat{D}^{BC}_{zz} \\
\hat{\Gamma}_{\phi\phi}= \hat{E}_{\phi \phi} \\
\hat{\Gamma}_{\rho z}=\hat{D}_{\rho z}+\hat{E}_{\rho z}^{BC}\\
\hat{\Gamma}_{z\rho}=\hat{D}_{z\rho}+\hat{E}_{z\rho}^{BC}.
\end{align}
Expressions for the operator matrices can be found in the appendix. 

\section{Micromagnetic Simulations}

Micromagnetic simulations were also performed using MuMax3 \cite{Vansteenkiste2014TheMuMax3} and compared with the results from the matrix method. The simulations were conducted by first relaxing a particular spin structure in zero magnetic field to obtain the ground state. Next, a small out-of-plane perturbation field of approximately 20~mT was applied, chosen such that the perturbation of the spins from the zero-field equilibrium state is a few percent. The spins were next allowed to precess after removing the perturbation field using a damping parameter of $\alpha=0.01$, and Fourier transforms of the $z$-component of the magnetization versus time were performed to obtain the spectra. Simulations with a sinusoidal driving field were conducted at selected resonance frequencies and the modes were constructed from the spin distributions saved out over two periods after a stead-state response to the driving field was reached, typically after approximately 50 oscillations. Simulations were conducted for both disks and rings in the vortex state, where the vortex state in the ring lacks the central out-of-plane core. Structures with $R$ and $L$ ranging from 100~to 350~nm and 1~to 10~nm, respectively, were considered.  Parameters typical for Permalloy were used for the magnetic layer: $M_s = 8\times10^5$~A/m, exchange of $A=1\times10^{-11}$~J/m$^2$, anisotropy was neglected, and the interfacial DMI was varied from 0 to 1.5~mJ/m$^2$, which is within the range of values that have been reported for heavy metal/ferromagnetic thin film bilayers such as Permalloy/Pt~\cite{Stashkevich2015ExperimentalInteraction,Nembach2015LinearFilms}. 

\section{Results and Discussion}

\begin{figure}[t]
	\includegraphics[width=\columnwidth]{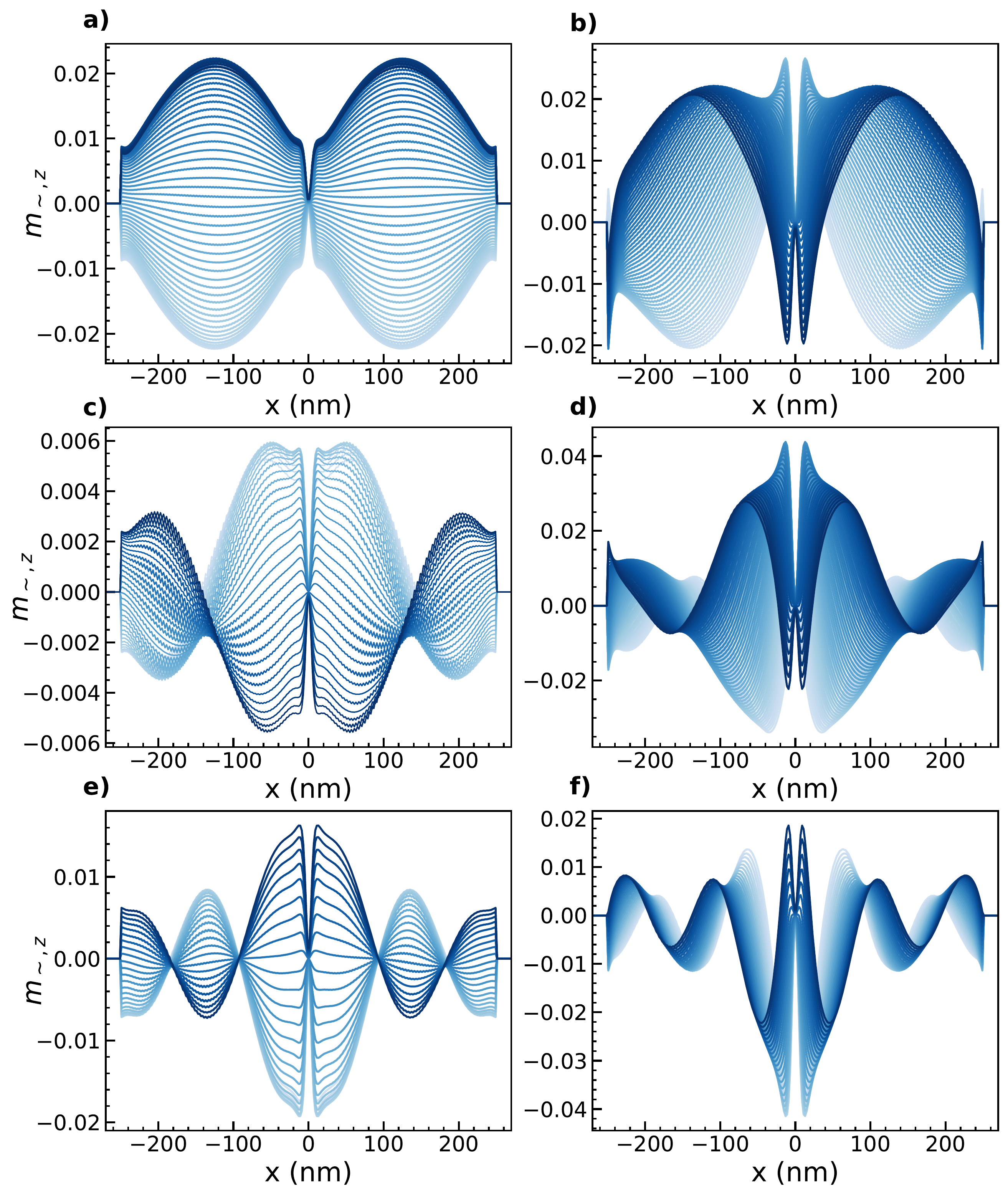}
    \caption{
    (Color online) Time-evolution of the cross sectional dynamic magnetization profiles $m_{\sim,z}(\rho)$ for the first three modes obtained from simulations with $R=250$~nm and $L=5$~nm. The left and right columns show the modes for $D = 0$ and 1.5~mJ/m$^2$, respectively. The light to dark lines are used to show the time evolution over one half period. The modes for $D=0$ are standing modes, whereas the modes for $D=1.5$~mJ/m$^2$ show less pronounced nodes and outward motion.
    }
    \label{fig:mz_profiles}
\end{figure}

\begin{figure}[t]
	\includegraphics[width=\columnwidth]{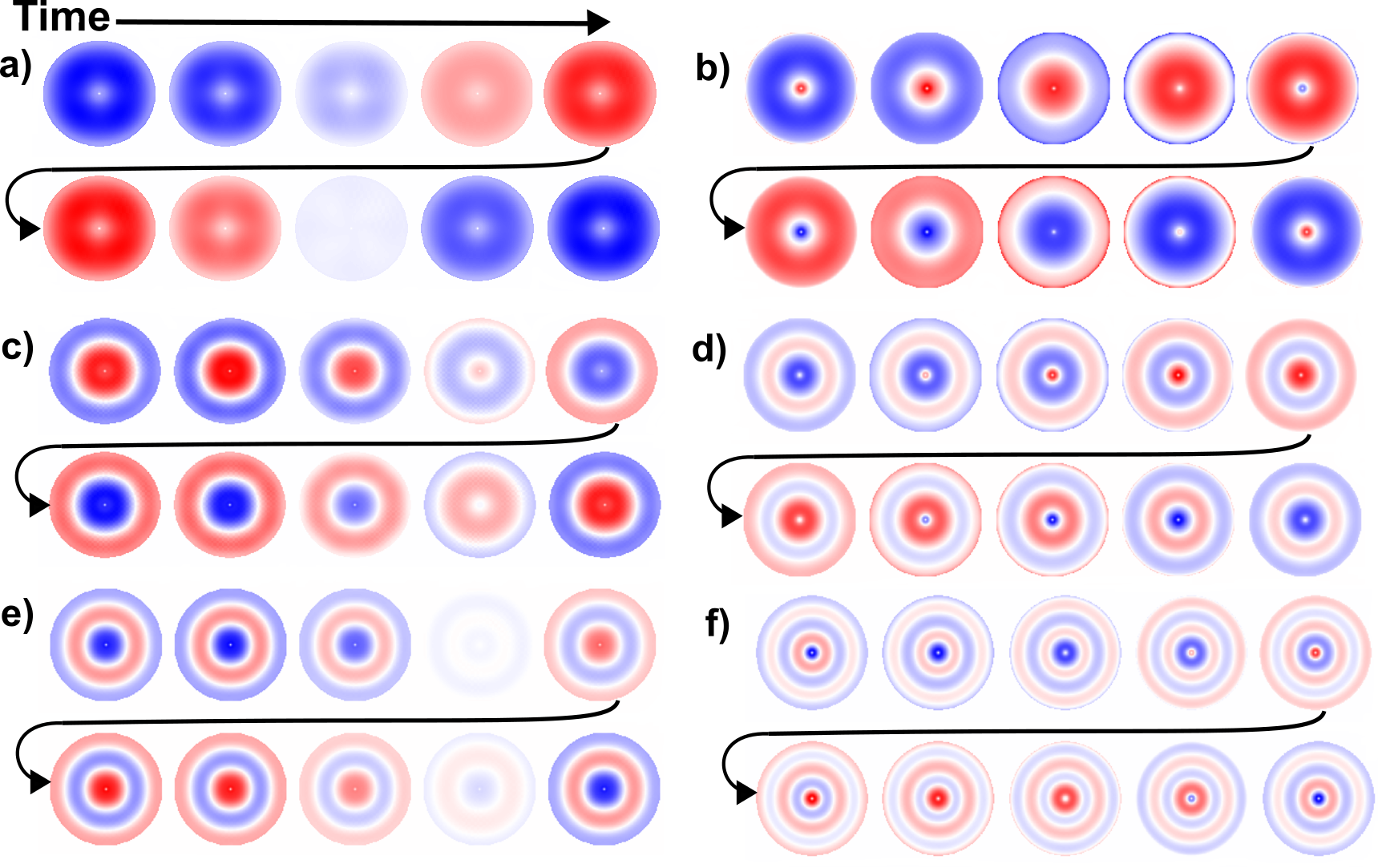}
    \caption{
    (Color online) Two dimensional m$_{\sim,z}(\rho)$ mode maps shown as a function of time for a disk of radius $R=250$~nm and $L=5$~nm over one period, obtained from simulations. The first, second and third modes for $D=0$ are shown in a), c), and e), and the first, second, and third modes for $D=1.5$~mJ/m$^2$ are shown in b), d) and f), respectively. Red, white, and blue represent positive, zero, and negative amplitudes, respectively. 
    }
    \label{fig:mz_modeMaps}
\end{figure}

\begin{figure*}[t!]
\centering
\hspace{0em}\includegraphics[scale=0.65,angle=0]{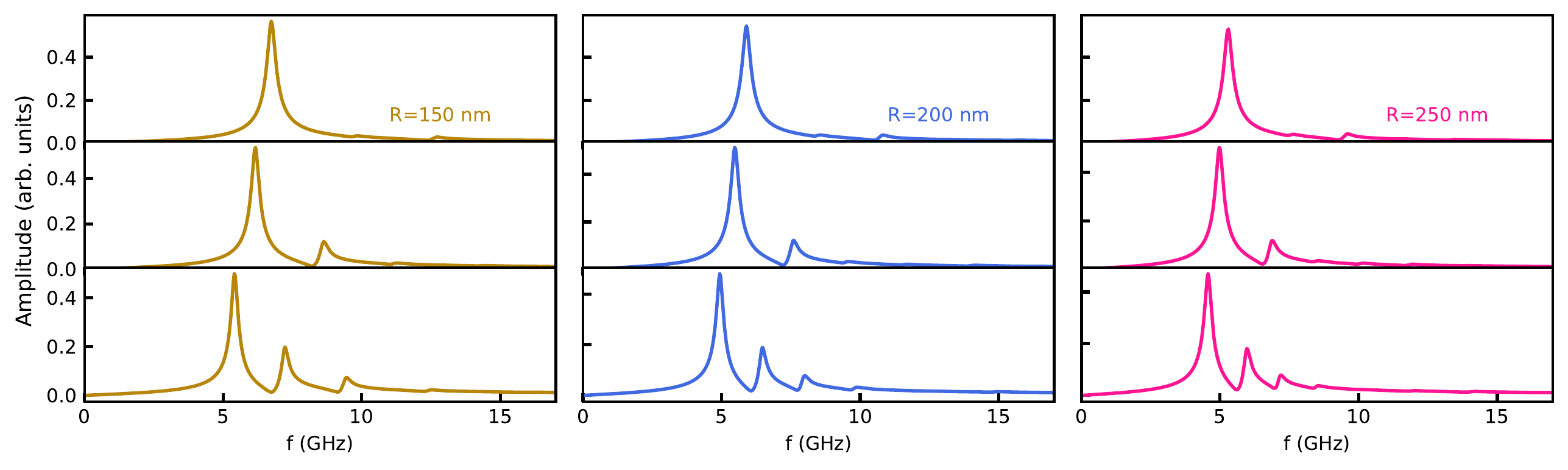}
\caption{Normalized spin wave spectra from simulations for disks with $L=$5~nm and $R$ from left to right of 150, 200, and 250~nm, where the $D$ values are, from top to bottom, 0~mJ/m$^2$, 1~mJ/m$^2$, and 1.5~mJ/m$^2$. The peaks correspond to radial-type vortex modes. Note that only down-shifted modes are observed for both vortex circulations/chiralities.}
\label{fig:spectra_R250L5}
\end{figure*} 

\begin{figure}[b]
	\includegraphics[width=\columnwidth]{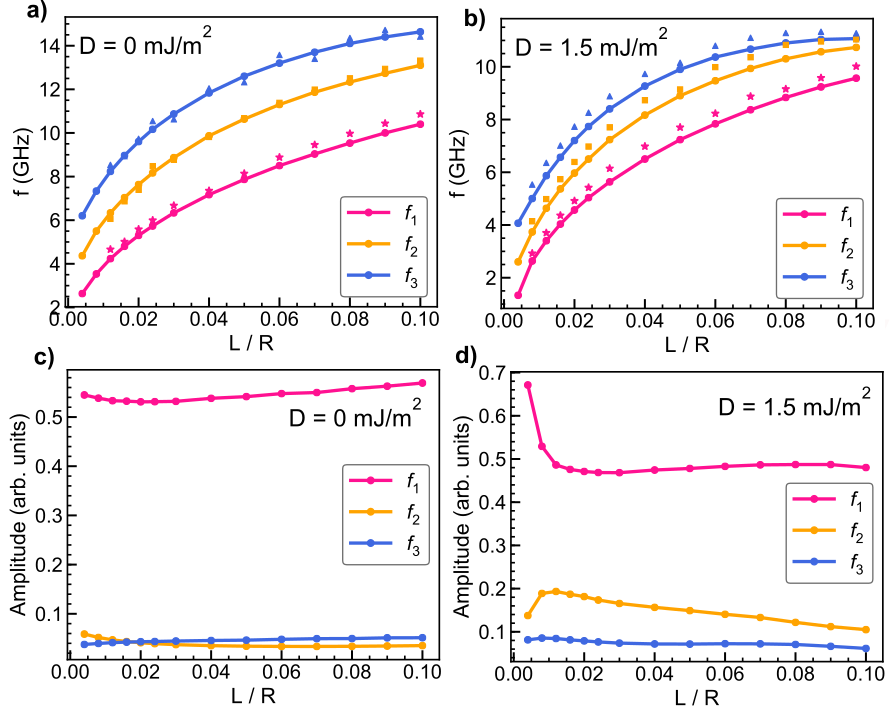}
    \caption{
    (Color online) Radial spin-wave mode eigenfrequencies as a function of the disk aspect ratio $L/R$ with $R=250$~nm for a) $D=0$~mJ/m$^2$, and b) $D=1.5$~mJ/m$^2$. The corresponding normalized spin wave amplitudes are shown in c) and d) for the same values of $D$. The dots show the full simulation results, while the closed symbols (*, $\square$, and $\triangle$ represent modes 1, 2, and 3, respectively) correspond to the real part of the semi-analytical solutions calculated using the LLD method. The lines connecting the frequency values from the simulations are guides to the eye.
    }
    \label{fig:f_A_vs_LR_R250}
\end{figure}

\begin{figure}[b]
	\includegraphics[width=\columnwidth]{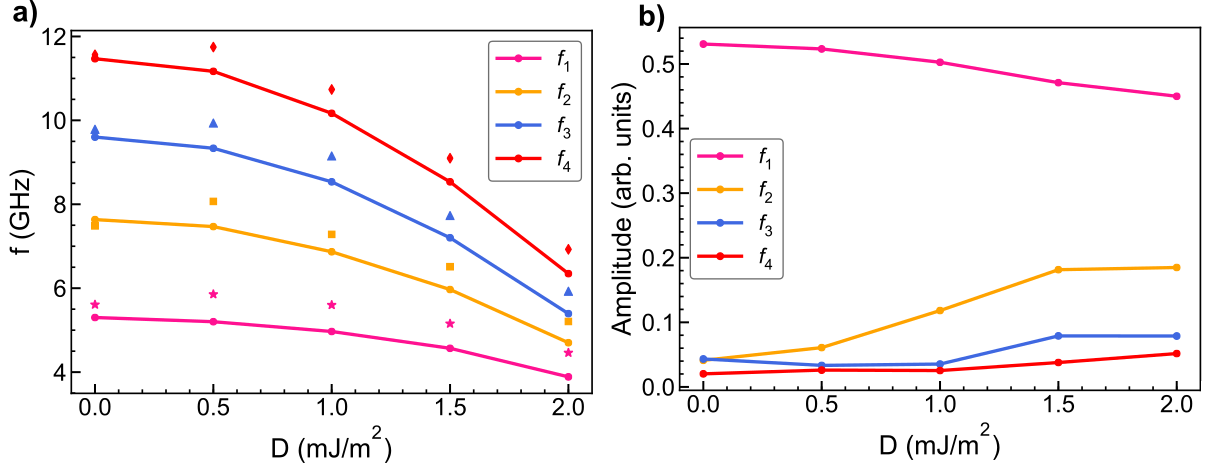}
    \caption{
    (Color online) Radial spin-wave mode a) eigenfrequencies and b) normalized mode amplitudes as a function of $D$ for a disk with $R=250$~nm and $L=5$~nm. The dots are the full simulation results while the closed symbols correspond to the real part of the semi-analytical LLD calculations. The lines connecting the simulation frequency values are guides to the eye.
    }
    \label{fig:f_A_vs_D_R250L5}
\end{figure}

The static equilibrium state and the dynamic modes were calculated using the 1D LLD approach and micromagnetic simulations. The tilt angles $\psi$ in Fig.~\ref{fig:vortex_tilt}, obtained using the 1D expressions, were used as input to calculate the the kernel $\hat{\Gamma}(\rho,\rho^{\prime})$ in Eqs.~(\ref{Eigenvealu-problem}) and~(\ref{Eigenvealu-problem-extended}). The static profiles obtained from the LLD approach and the micromagnetic simulations agree well, as shown in Fig.~\ref{fig:vortex_tilt}b. The DMI leads to an in-plane tilt of the magnetization that increases with increasing $D$. The DMI also leads to changes in $m_{0,z}$, as shown in Fig.~\ref{fig:COREmz}. For a given $L$ and $R$, the core size increases with increasing $|D|$, where the size depends on the magnitude but not the sign of $D$, and the spins near the edges tilt out-of-plane in the direction opposite to the core magnetization. Both of these effects are most pronounced for small $L$; for $L = 10$, 5, and 1~nm, the core sizes at $D = 2$~mJ/m$^2$ are 28\%, 45\%, and 100\% larger than the core at $D=0$, respectively. Note that these calculations assume a constant $D$ to to provide a straightforward means to compare the effects of DMI with that of $L$, which mainly changes the demagnetization energy, but since the effect is interfacial in nature, the DMI should be very small for $L = 10$~nm in a real sample. The tilting of the edges is also observed in rings and is a general feature of patterned magnetic structures with DMI. 

The dynamic modes of the magnetic vortex are quantized along the radial direction. Figure~\ref{fig:mzmrEigenvector} shows a comparison of the eigenvectors obtained from the LLD method and the simulations for the first three modes and the two agree well. These modes are also shown as cross-sectional plots and full mode maps of the out-of-plane motion $m_{\sim,z}$ in Figs.~\ref{fig:mz_profiles} and~\ref{fig:mz_modeMaps}, respectively. For $D=0$ (left panels), the modes are standing wave excitations with well defined nodes and antinodes, whereas for $D=1.5$~mJ/m$^2$ (right panels), the modes have similar wavelengths but are propagating rather than standing waves. Nodes are still present but they are much less pronounced than they are for the $D=0$ case. Similar effects have been noticed in confined 1-D systems with DMI \cite{Zingsem2019UnusualWaves}. As shown in Fig.~\ref{fig:mz_modeMaps}, the spin wave excitations appear to move from the center, outwards. The edge amplitudes are also much larger when DMI is present. Cross sections and full mode maps of the in-plane dynamic magnetization $\widetilde{m}_{\sim,s}$ show similar behavior (see appendix). 

The direction of the propagation depends on the circulation of the vortex $c$ and the sign of $D$, but it is independent of the polarity $p$ of the vortex. The propagation direction is inwards for $cD<1$, and outward for $cD>1$, where $c=1$ ($-1$) corresponds to a counterclockwise (clockwise) circulation. These same effects are observed in ring-shaped structures in a vortex state that lack the complication of the central out-of-plane core and the resulting DMI-induced tilt angle $\psi$, hence the propagating nature of the modes is due to the DMI and is not a consequence of the static tilt. In fact, the sign of the static tilt depends on the sign of $D$ and the polarity $p$ of the vortex core but is independent of $c$. An inward (outward) static tilt is observed for $Dp>1$ ($Dp<1$) where $p=1$ corresponds to a polarity in the $+\hat{z}$ direction. The resonance frequencies are the same for any $D$, $p$, or $c$. 

The DMI magnitude has a significant effect on the resonant frequencies and mode amplitudes. Fig.~\ref{fig:spectra_R250L5} shows spectra obtained from micromagnetic simulations for selected $R$ and $D$. The spectra for $D=0$ show a strong mode that is the lowest-order radial mode, and a weaker mode that is the third order mode. A small peak that is due to the second mode is barely visible between the first and third modes. The spectra for $D>0$, in contrast, show strong peaks for each of the first, second, and third order modes. Without DMI the even modes are only weakly excited because they have a net moment near zero and the odd modes are favored, whereas with DMI, the odd and even modes have comparable amplitudes. A similar effect was observed in Ref.~\onlinecite{Zingsem2019UnusualWaves} for simulations of saturated elements. The DMI leads to a reduction in symmetry that allows all possible modes to couple to a uniform driving microwave field. Fig.~\ref{fig:f_A_vs_LR_R250} shows that the frequency increases as a function of $L/R$ for $D=1.5$~mJ/m$^2$ (panel b) in a similar manner as what is observed for $D=0$ (panel a). The LLD results for $D=0$ are the same as those calculated using Ref.~[\onlinecite{Guslienko2005Vortex-stateDots}] if the exchange contribution is neglected. Here the exchange is included, which slightly raises the calculated resonance frequencies. Fig.~\ref{fig:f_A_vs_D_R250L5} shows that the frequencies of all of the observed modes decrease as a function of increasing $D$ for $R=250$~nm and $L=5$~nm, and similar trends are observed for other structure dimensions (additional plots are included in the appendix). In all cases the LLD and micromagnetic simulation results agree well. 

The spectra for the vortices with the DMI show shifts in the frequency of the respective modes that are similar to what is observed for spin waves with comparable wavelengths in extended films with one important difference: the frequency shifts in the confined structures are always to lower frequencies as compared to what is observed at $D=0$, and this corresponds to a dominant mode direction.  For a thin film, surface waves with a given wavelength that propagate in opposite directions have different frequencies, and surface spin waves at a particular frequency that travel in opposing directions have different wavelengths. In a confined structure, the wavelength quantization imposed by the structure size and/or the spin texture creates a situation where it should be possible to excite just one of the inward or outward propagating modes of a particular order at a slightly smaller frequency than the $D=0$ resonance frequency, and the other at a slightly higher frequency. The outward-propagating mode is shown in Fig.~\ref{fig:mz_modeMaps}; as mentioned previously, the direction can be changed by either reversing the sign of $D$ or the vortex chirality. The oppositely directed, higher frequency mode is, however, suppressed in the simulations and, according to the LLD calculations this is because this mode is not an eigenmode of the system. This aspect of how the DMI affects spin dynamics in confined geometry could be useful for enhancing non-reciprocity for magnonics applications. 

In Fig.~\ref{fig:f_A_vs_LR_R250}, the resonance frequencies shown from the LLD calculations are the real part of $\omega$, as calculated from Eq.~(\ref{Eigenvealu-problem-extended}). For $D=0$, the eigenvalues are real, but for $|D|>0$ the eigenvalues are complex. For $D=0$, small imaginary parts of order $~1\times10^{-14}$ are obtained, which is a numerical artifact; the imaginary parts of the eigenfrequencies of the first mode are as large as $\sim1$~percent for $|D|>0$, which may be indicative of a DMI contribution to the damping. In the micromagnetic simulations the linewidths of the first mode are slightly narrower in Fig.~\ref{fig:spectra_R250L5} for larger $D$. The linewidths $\Delta f$ are 0.47 and 0.40~GHz for $R=250$~nm with $D=0$ and 1.5~mJ/m$^2$, respectively, which correspond to fractional linewidths of $\Delta f/f_1$ of 8.81\% and 8.75\%, respectively. Theoretical calculations\cite{Moon2013Spin-waveInteraction} predict that the DMI should lead to a change in not just the frequency but also the linewidth for extended thin films, so it is not surprising that the DMI also lead to linewidth changes for modes in confined structures. 

\section{Conclusions}
In conclusion, the DMI lead to important modifications of the \textit{static spin state} and the \textit{dynamic excitations} of magnetic vortices. These modifications provide insight into the types of effects that should be observed for other patterned magnetic structures. These changes are captured well by the LLD method, which provides a rapid means to gain insight into the behaviour not just for vortex-based systems, but also for other spin textures with cylindrical symmetry. For the vortex, the inclusion of the DMI induces a static in-plane tilt of the spins that increases as a function of $D$ and that disappears if the core is absent, i.e., in a magnetic ring. In the presence of DMI the radial-type vortex modes are still quantized, however, the mode frequencies are shifted, the modes are propagating rather than standing modes, and the even and odd modes are both excited effectively by a spatially uniform mode due to the symmetry breaking provided by the DMI. There are also differences of at the edges, specifically an out-of-plane static tilt at the edges is observed, and the edges show higher dynamic amplitudes in the presence of the DMI. The changes in the mode frequencies induced by the DMI are similar to what is predicted for extended films, however, unlike the case of an extended film, only the downward-shifted modes are eigenmodes of the dynamic equations, which suggests that the combination of DMI and confined geometries could lead to new strategies for non-reciprocal spin wave excitation, or, in the case of vortices, the outward-only mode could serve as a point-like source for spin waves if coupled to an extended film.

\begin{acknowledgments}
We acknowledge helpful discussions with Robert Camley about the DMI. This work is supported by National Science Foundation DMR Grant Number 1709525. JD and KLL were supported by a UCCS CREW Award.
\end{acknowledgments}

\appendix*
\section{Definitions of matrix operators}
\begin{figure}[h]
	\includegraphics[width=\columnwidth]{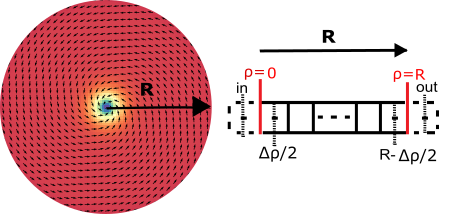}
    \caption{(Color online) Illustration of the discretization scheme used for the LLD calculations. 
    }
    \label{fig:dmi_BC}
\end{figure}
The differential DMI and exchange operators can be written in matrix form using the finite difference method. We discretize the first and second order derivatives using a central difference scheme at each mesh site. For example the central difference for the first and second order derivatives of $m_{\rho}$ at the radial position $\rho$ are 
\begin{align}
\label{finite-diff}
\frac{dm_{\rho}}{d \rho} \approx \frac{-m_{\rho}(\rho-\Delta\rho)+m_{\rho}(\rho+\Delta\rho)}{2 \Delta \rho} \\
\frac{d^2m_{\rho}}{d \rho^2} \approx \frac{m_{\rho}(\rho-\Delta\rho)-2m_{\rho}(\rho)+m_{\rho}(\rho+\Delta\rho)}{ (\Delta \rho)^2}
\end{align}
where $m_{\rho}(\rho \pm \Delta\rho)$ are the nearest neighbor magnetic moments of $m_{\rho}(\rho)$ at mesh site $\rho$, and $\Delta \rho$ is the mesh discretization in the $\rho$ direction. To obtain the derivatives for the outermost cell at $\rho=R-\Delta\rho/2$, we extrapolate to find the hypothetical magnetization at the cell $m_{out,\rho}$, as shown in Fig.~\ref{fig:dmi_BC}, and use this value to compute the first and second order derivatives of the magnetization. The missing value $m_{out,i}$, with $i=\rho,\phi$, or $z$, is extrapolated as follows
\begin{align}
m_{out,i} \approx m_{i}(R-\Delta \rho/2) +  \Delta \rho \left.\frac{dm_i}{d\rho} \right|_{\rho=R} 
\end{align}
where the derivative of the magnetization at the boundary $\rho=R$ is obtained by considering the DMI boundary conditions, Eq.~(\ref{dmi-BC}). Therefore, the outer missing neighbors are
\begin{multline}
m_{out,\rho} \approx m_{\rho}(R-\Delta\rho/2) +  \Delta \rho  \left.\frac{dm_\rho}{d\rho} \right|_{\rho=R}=  \\
m_{\rho}(R-\Delta\rho/2) - \Delta \rho \frac{D  m_{z}(R-\Delta\rho/2)}{2A_{ex}}
\end{multline}

\begin{multline}
m_{out,z} \approx m_{z}(R-\Delta\rho/2) +  \Delta \rho  \left.\frac{dm_z}{d\rho} \right|_{\rho=R}=  \\
m_{z}(R-\Delta\rho/2) + \Delta \rho \frac{D  m_{\rho}(R-\Delta\rho/2)}{2A_{ex}}
\end{multline}

\begin{multline}
m_{out,\phi} \approx m_{\phi}(R-\Delta\rho/2) +  \Delta \rho  \left.\frac{dm_{\phi}}{d\rho} \right|_{\rho=R}=  \\
m_{\phi}(R-\Delta\rho/2) 
\end{multline}
Moreover, the magnetic moment at the cell $m_{in,i}$ with $i=\rho,\phi$, or $z$, as Fig.~\ref{fig:dmi_BC} shows, is obtained considering the symmetry in the spin distribution of the vortex state with DMI. In this manner, the in-plane magnetization at the cell site $m_{in,\rho}$ or $m_{in,\phi}$ is obtained by flipping the sign of the corresponding magnetization at the position $\rho= \Delta \rho /2$ while the out-of-plane magnetization $m_{in,z}$ preserves its direction. In both cases the magnitude of the magnetization is preserved.
\begin{align}
m_{in,\rho} = -m_{\rho}(\Delta\rho/2) \\
m_{in,\phi} = -m_{\phi}(\Delta\rho/2) \\
m_{in,z} = m_{z}(\Delta\rho/2)   
\end{align}
The first and last rows of the DMI and exchange matrices below are constructed using the presented symmetry considerations and DMI boundary conditions. Note that the matrices $\hat{D}_{\rho\rho}^{BC}$, $\hat{D}_{zz}^{BC}$, $\hat{E}_{\rho z}^{BC}$ and
$\hat{E}_{z \rho}^{BC}$ contain only terms due to DMI boundary conditions, while the remaining terms are incorporated into the $\hat{D}_{\rho z}$, etc. matrices. 

Using the following notation
\begin{gather}
\begingroup 
\renewcommand{\arraystretch}{0.5}
\setlength\arraycolsep{2pt}
 \hat{\mathbb{P}} = 
\begin{pmatrix} 
\rho_1 &   &    &    &         &       &    &    &   \\
 &  \rho_2  &   &    &         &       &    &    &   \\
   &  & \rho_3  &   &         &       &    &    &   \\
   &    &    &    & \ddots  &       &    &    &    \\
   &    &    &    &         &     &  \rho_{n-2} &   &    \\
   &    &    &    &         &       &  & \rho_{n-1}  &   \\
   &    &    &    &         &       &    &  & \rho_{n}
\end{pmatrix},
\endgroup
\end{gather}
and
\begin{gather}
\begingroup 
\renewcommand{\arraystretch}{0.5}
\setlength\arraycolsep{2pt}
 \hat{F}^{BC }= 
\begin{pmatrix} 
0 &   &    &    &         &       &    &    &   \\
 &  0 &   &    &         &       &    &    &   \\
   &  & 0  &   &         &       &    &    &   \\
   &    &    &    & \ddots  &       &    &    &    \\
   &    &    &    &         &     &  0 &   &    \\
   &    &    &    &         &       &  & 0 &   \\
   &    &    &    &         &       &    &  & \frac{D \Delta \rho}{2A_{ex}} 
\end{pmatrix},
\endgroup
\end{gather}
the matrix forms of the DMI and exchange differential operators are
\begin{multline}
\begingroup 
\renewcommand{\arraystretch}{0.5}
\setlength\arraycolsep{2pt}
\hat{D}_{\rho z}\approx 
\frac{2D}{\mu_0 M_s^2} 
\frac{1}{ 2 \Delta \rho}
\begin{pmatrix} 
-1 & 1 &    &    &         &       &    &    &   \\
-1 &  0 & 1  &    &         &       &    &    &   \\
   & -1 & 0  & 1  &         &       &    &    &   \\
   &    &    &    & \ddots  &       &    &    &    \\
   &    &    &    &         & -1    &  0 & 1  &    \\
   &    &    &    &         &       & -1 & 0  & 1  \\
   &    &    &    &         &       &    & -1 & 1 
\end{pmatrix},
\endgroup
\end{multline}

\begin{multline}
\begingroup 
\renewcommand{\arraystretch}{0.5}
\setlength\arraycolsep{2pt}
\hat{D}_{z \rho} \approx 
-\frac{2D}{\mu_0 M_s^2} 
\frac{1}{ 2 \Delta \rho}
\begin{pmatrix} 
1 & 1 &    &    &         &       &    &    &   \\
-1 &  0 & 1  &    &         &       &    &    &   \\
   & -1 & 0  & 1  &         &       &    &    &   \\
   &    &    &    & \ddots  &       &    &    &    \\
   &    &    &    &         & -1    &  0 & 1  &    \\
   &    &    &    &         &       & -1 & 0  & 1  \\
   &    &    &    &         &       &    & -1 & 1 
\end{pmatrix} 
\endgroup
\\ {}-\frac{2D}{\mu_0 M_s^2}\hat{\mathbb{P}}^{-1},
\end{multline}

\begin{align}
\hat{D}_{\rho \rho }= \hat{D}_{zz }\approx
\frac{2D}{\mu_0 M_s^2} \frac{1}{2\Delta \rho} \hat{F}^{BC},
\end{align}

\begin{multline}
\begingroup 
\renewcommand{\arraystretch}{0.5}
\setlength\arraycolsep{2pt}
\hat{E}_{\rho \rho}\approx
\frac{2 A_{ex}}{\mu_0 M_s^2} \left[
\frac{1}{ (\Delta \rho)^2 }
\begin{pmatrix} 
-3 & 1 &    &    &         &       &    &    &   \\
 1 &  -2 & 1  &    &         &       &    &    &   \\
   & 1 & -2  & 1  &         &       &    &    &   \\
   &    &    &    & \ddots  &       &    &    &    \\
   &    &    &    &         & 1    &  -2 & 1  &    \\
   &    &    &    &         &       & 1 & -2  & 1  \\
   &    &    &    &         &       &    & 1 & -1 
\end{pmatrix} \right.
\endgroup
\\ {}+  \left. \frac{1}{2 \Delta \rho} \hat{\mathbb{P}}^{-1}
\begingroup 
\renewcommand{\arraystretch}{0.5}
\setlength\arraycolsep{2pt}
\begin{pmatrix} 
1 & 1 &    &    &         &       &    &    &   \\
-1 &  0 & 1  &    &         &       &    &    &   \\
   & -1 & 0  & 1  &         &       &    &    &   \\
   &    &    &    & \ddots  &       &    &    &    \\
   &    &    &    &         & -1    &  0 & 1  &    \\
   &    &    &    &         &       & -1 & 0  & 1  \\
   &    &    &    &         &       &    & -1 & 1 
\end{pmatrix} 
\endgroup 
-\hat{\mathbb{P}}^{-2}
\right],
\end{multline}

\begin{align}
\hat{E}_{\phi \phi} = \hat{E}_{\rho \rho},
\end{align}

\begin{multline}
\begingroup 
\renewcommand{\arraystretch}{0.5}
\setlength\arraycolsep{2pt}
\hat{E}_{z z} \approx
\frac{2 A_{ex}}{\mu_0 M_s^2} \left[
\frac{1}{ (\Delta \rho)^2 }
\begin{pmatrix} 
-1 & 1 &    &    &         &       &    &    &   \\
 1 &  -2 & 1  &    &         &       &    &    &   \\
   & 1 & -2  & 1  &         &       &    &    &   \\
   &    &    &    & \ddots  &       &    &    &    \\
   &    &    &    &         & 1    &  -2 & 1  &    \\
   &    &    &    &         &       & 1 & -2  & 1  \\
   &    &    &    &         &       &    & 1 & -1 
\end{pmatrix} \right.
\endgroup
\\ {}+  \frac{1}{2 \Delta \rho} \hat{\mathbb{P}}^{-1}
\begingroup 
\renewcommand{\arraystretch}{0.5}
\setlength\arraycolsep{2pt}
\left.
\begin{pmatrix} 
-1 & 1 &    &    &         &       &    &    &   \\
-1 &  0 & 1  &    &         &       &    &    &   \\
   & -1 & 0  & 1  &         &       &    &    &   \\
   &    &    &    & \ddots  &       &    &    &    \\
   &    &    &    &         & -1    &  0 & 1  &    \\
   &    &    &    &         &       & -1 & 0  & 1  \\
   &    &    &    &         &       &    & -1 & 1 
\end{pmatrix} \right],
\endgroup
\end{multline}

\begin{multline}
\hat{E}_{\rho z}= -\hat{E}_{z \rho} \approx
-\frac{2 A_{ex}}{\mu_0 M_s^2} \left[
\frac{1}{ ( \Delta \rho)^2}
\hat{F}^{BC} \right.\\ +
\left.
\frac{1}{2 \Delta \rho} \mathbb{P}^{-1} \hat{F}^{BC} \right].
\end{multline}

The integral demagnetization operators, Eqs.(\ref{Arr}-\ref{Azz}), are, in matrix form,
\begin{widetext}
\begin{multline}
\hat{A}_{\rho\rho} = 
\begingroup 
\renewcommand{\arraystretch}{0.9}
\setlength\arraycolsep{2pt}
\begin{pmatrix} 
g_{\rho\rho}(\rho_1,\rho'_1) & g_{\rho\rho}(\rho_1,\rho'_2)   &  g_{\rho\rho}(\rho_1,\rho'_3)   & \cdots  &         &       &    &    &   \\
  g_{\rho\rho}(\rho_2,\rho'_1)     &  g_{\rho\rho}(\rho_2,\rho'_2) &  g_{\rho\rho}(\rho_2,\rho'_3)  &  \cdots  &         &       &    &    &   \\
 g_{\rho\rho}(\rho_3,\rho'_1)   & g_{\rho\rho}(\rho_3,\rho'_2)  & g_{\rho\rho}(\rho_3,\rho'_3)   & \cdots  &         &    &    &    &   \\
      \vdots    &    \vdots   &    \vdots   &    &   &       &    &    &    \\
  g_{\rho\rho}(\rho_{n-2},\rho'_1)  &   g_{\rho\rho}(\rho_{n-2},\rho'_2)  &    g_{\rho\rho}(\rho_{n-2},\rho'_3) &  \cdots  &         &     &   &   &    \\
  g_{\rho\rho}(\rho_{n-1},\rho'_1)&  g_{\rho\rho}(\rho_{n-1},\rho'_2)  &  g_{\rho\rho}(\rho_{n-1},\rho'_3)  &  \cdots  &         &       &  & \ &   \\
  g_{\rho\rho}(\rho_{n},\rho'_1)&  g_{\rho\rho}(\rho_{n},\rho'_2)  &  g_{\rho\rho}(\rho_{n},\rho'_3)  &  \cdots  &         &       &    &  & 
\end{pmatrix} 
\begin{pmatrix} 
\rho'_1 &   &    &    &         &       &    &    &   \\
      &  \rho'_2 &   &    &         &       &    &    &   \\
   &  & \rho'_3  &   &         &       &    &    &   \\
       &    &    &    & \ddots  &       &    &    &    \\
   &    &    &    &         &     &  \rho'_{n-2} &   &    \\
  &    &    &    &         &       &  & \rho'_{n-1} &   \\
  &    &    &    &         &       &    &  & \rho'_n 
\end{pmatrix} 
\endgroup \Delta \rho',
\end{multline}

\begin{multline}
\hat{A}_{zz} = 
\begingroup 
\renewcommand{\arraystretch}{0.9}
\setlength\arraycolsep{2pt}
\begin{pmatrix} 
g_{zz}(\rho_1,\rho'_1) & g_{zz}(\rho_1,\rho'_2)   &  g_{zz}(\rho_1,\rho'_3)   & \cdots  &         &       &    &    &   \\
  g_{zz}(\rho_2,\rho'_1)     &  g_{zz}(\rho_2,\rho'_2) &  g_{zz}(\rho_2,\rho'_3)  &  \cdots  &         &       &    &    &   \\
 g_{zz}(\rho_3,\rho'_1)   & g_{zz}(\rho_3,\rho'_2)  & g_{zz}(\rho_3,\rho'_3)   & \cdots  &         &    &    &    &   \\
      \vdots    &    \vdots   &    \vdots   &    &   &       &    &    &    \\
  g_{zz}(\rho_{n-2},\rho'_1)  &   g_{zz}(\rho_{n-2},\rho'_2)  &    g_{zz}(\rho_{n-2},\rho'_3) &  \cdots  &         &     &   &   &    \\
  g_{zz}(\rho_{n-1},\rho'_1)&  g_{zz}(\rho_{n-1},\rho'_2)  &  g_{zz}(\rho_{n-1},\rho'_3)  &  \cdots  &         &       &  & \ &   \\
  g_{zz}(\rho_{n},\rho'_1)&  g_{zz}(\rho_{n},\rho'_2)  &  g_{zz}(\rho_{n},\rho'_3)  &  \cdots  &         &       &    &  & 
\end{pmatrix} 
\begin{pmatrix} 
\rho'_1 &   &    &    &         &       &    &    &   \\
      &  \rho'_2 &   &    &         &       &    &    &   \\
   &  & \rho'_3  &   &         &       &    &    &   \\
       &    &    &    & \ddots  &       &    &    &    \\
   &    &    &    &         &     &  \rho'_{n-2} &   &    \\
  &    &    &    &         &       &  & \rho'_{n-1} &   \\
  &    &    &    &         &       &    &  & \rho'_n 
\end{pmatrix} 
\endgroup \Delta \rho',
\end{multline}
\end{widetext}
\noindent
where the magnetostatic kernels $g_{\rho\rho}$ and $g_{zz}$ should be written in matrix form with $\rho$ as columns, and $\rho'$ as rows following Eqs.~(\ref{grdemag}) and~(\ref{gzdemag}). Note that these kernels have numerically integrable singularities at $\rho=\rho'$ since the elliptic function $K_{ell}(\gamma^2)$ goes to infinity at $\rho=\rho'$. Due to the singularities, and because $g_{\rho\rho}$ changes rapidly near $\rho=\rho'$, the diagonal and near-diagonal terms of the demagnetization kernels were numerically integrated over each cell to obtain the average value within the cell $\bar{A}_{\rho\rho,ij}=\frac{1}{\rho'_j \Delta \rho'} \int_{\rho'_j - \Delta \rho'/2}^{\rho'_j+\Delta \rho'/2} g_{\rho\rho}(\rho_i,\rho')\rho' d\rho'$. The effective fields calculated using these operators agree well with the effective fields calculated using MuMax3.

\section{Dynamic $\widetilde{m}_{\sim,s}(\rho)$ profiles and results for additional aspect ratios}

\begin{figure}[h]
	\includegraphics[width=\columnwidth]{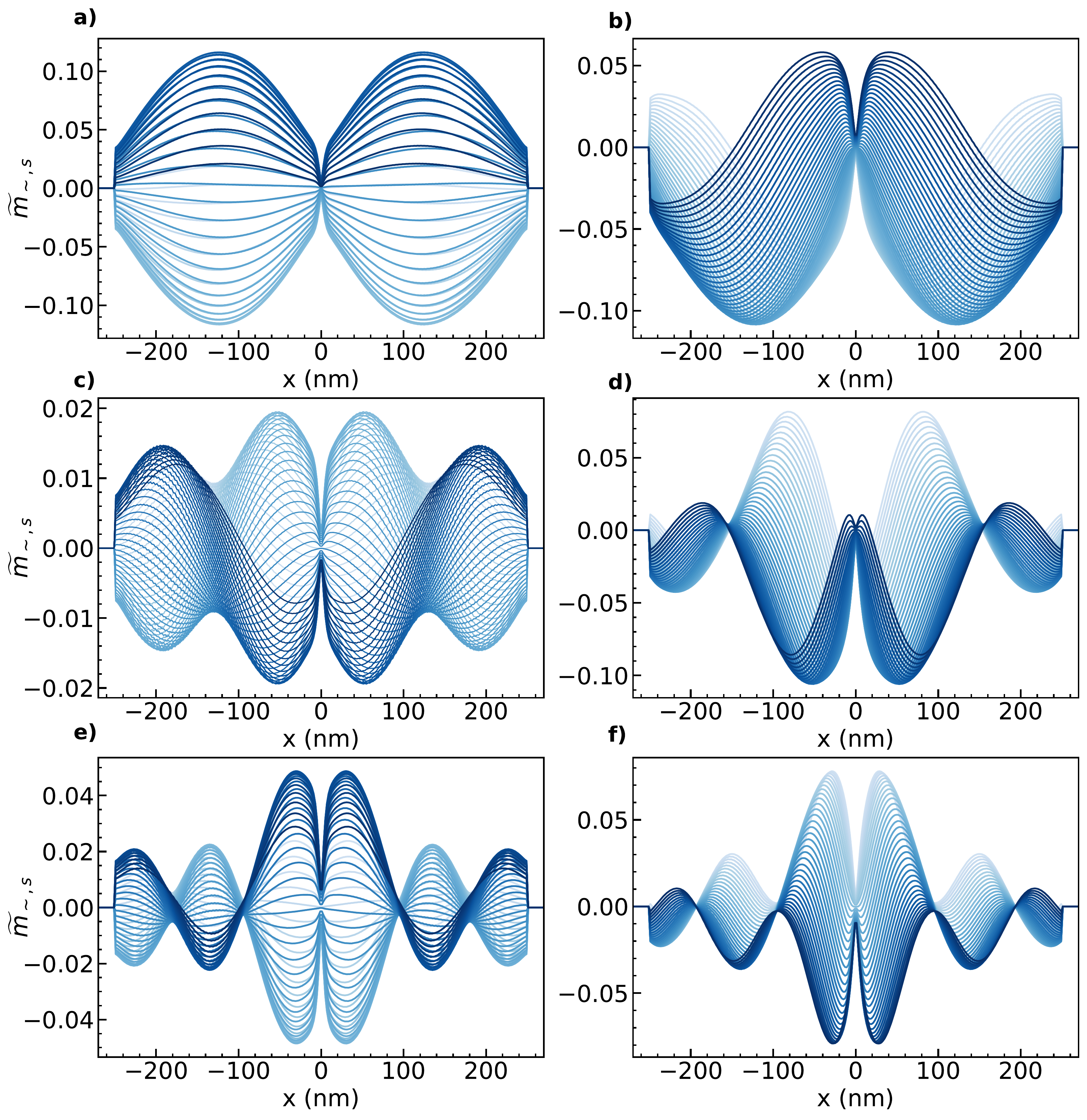}
    \caption{(Color online) Evolution of the cross sectional dynamic magnetization profiles $\widetilde{m}_{\sim,s}(\rho)$ for the same modes as Fig.~\ref{fig:mz_profiles} with $R=250$~nm and $L=5$~nm. The right and left columns show the modes for $D = 0$ and 1.5~mJ/m$^2$, respectively, for the first three modes. 
    }
    \label{fig:mr_profiles}
\end{figure}

\begin{figure}[h]
	\includegraphics[width=\columnwidth]{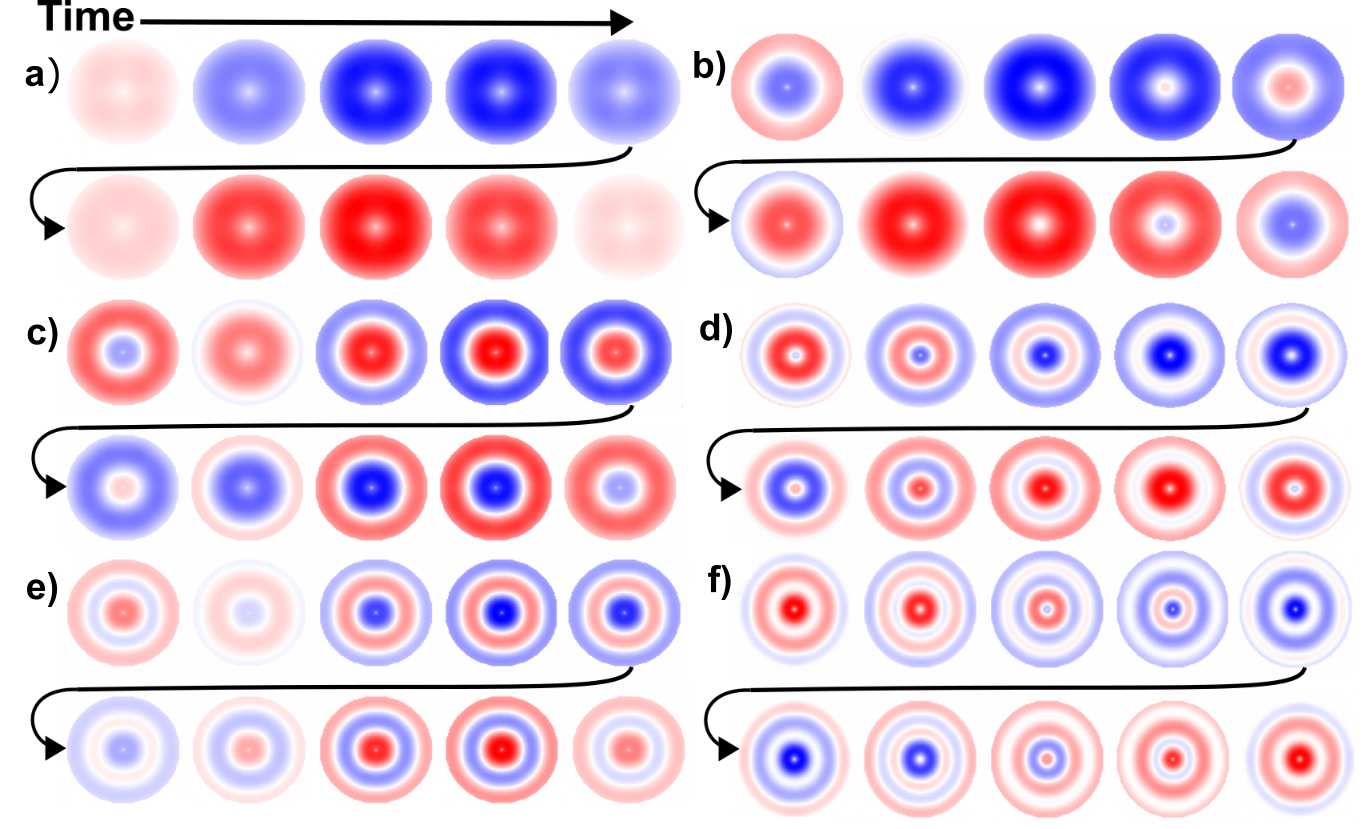}
    \caption{(Color online) Two dimensional $\widetilde{m}_{\sim,s}(\rho)$ mode maps shown as a function of time for a disk of radius $R=250$~nm and $L=5$~nm over one period for the same modes as Fig.~\ref{fig:mz_modeMaps}. The first, second and third modes for $D=0$ are shown in a), c), and e), and the first, second, and third modes for $D=$1.5~mJ/m$^2$ are shown in b), d) and f), respectively. Red, white, and blue represent positive, zero, and negative amplitudes, respectively. 
    }
    \label{fig:mr_modeMaps}
\end{figure}

The $\widetilde{m}_{\sim,s}(\rho)$ cross sectional and full dynamic mode profiles are shown in Figs.~\ref{fig:mr_profiles} and~\ref{fig:mr_modeMaps}, respectively, for the same parameters used in Figs.~\ref{fig:mz_profiles} and~\ref{fig:mz_modeMaps} ($R=$250~nm and $L=$5~nm) for $D = 0$ and 1.5~mJ/m$^2$. The $\widetilde{m}_{\sim,s}(\rho)$ and $m_{\sim,z}(\rho)$ profiles show similar characteristics. Note that the nodes for the second mode with $D=0$ shown in Fig.~\ref{fig:mr_profiles} are not as well defined as they are for the $m_{\sim,z}(\rho)$ profiles in Fig.~\ref{fig:mz_profiles}, which is likely because the second mode for $D=0$ is only weakly excited by the spatially uniform excitation field that was used for these simulations. 

\begin{figure}[b!]
	\includegraphics[width=\columnwidth]{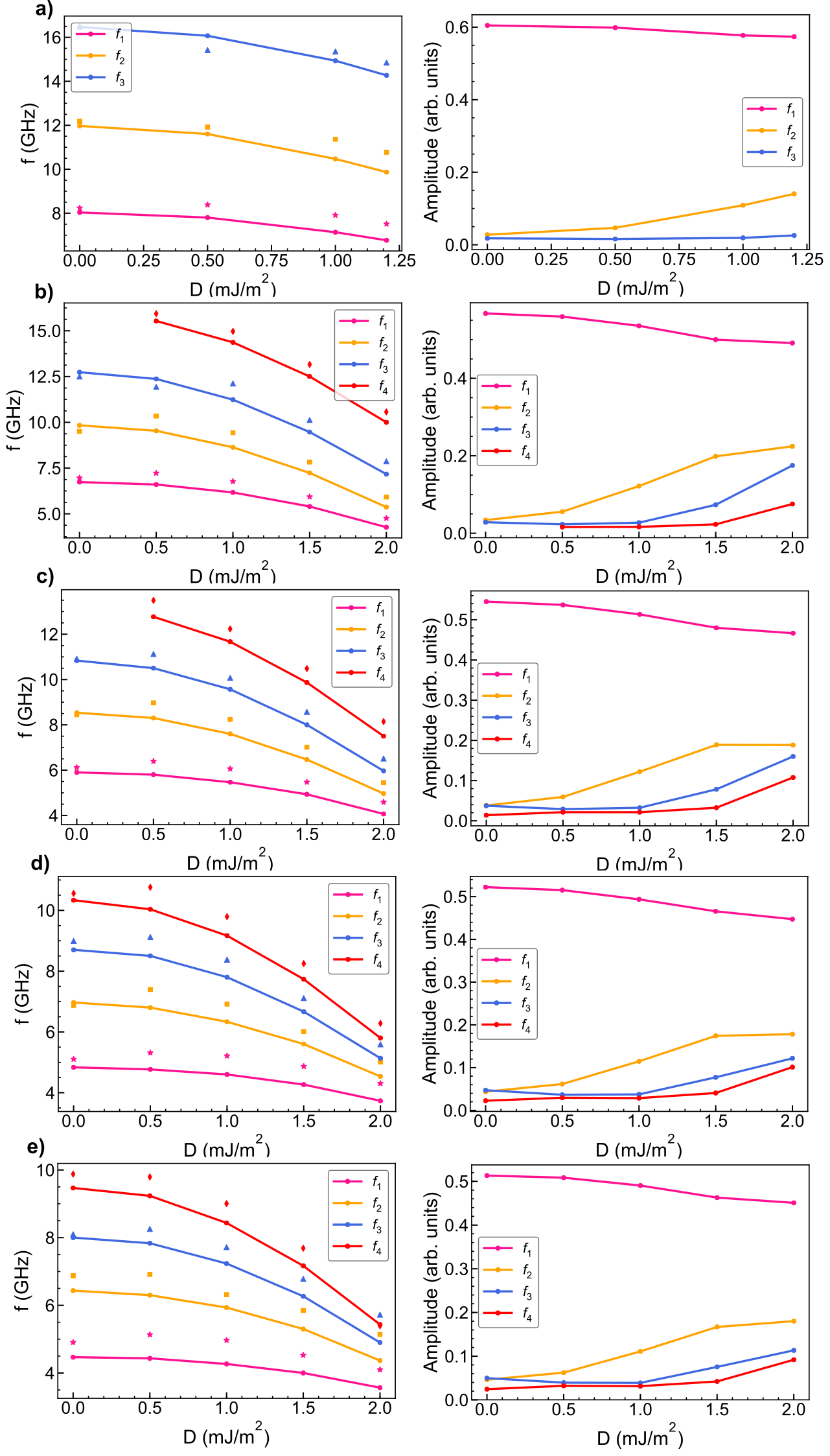}
    \caption{
    (Color online) Radial spin wave mode eigenfrequencies and normalized radial spin wave amplitudes as function of $D$ for a) $R=$100~nm, b) $R=$150~nm, c) $R=$200~nm, d) $R=$300~nm, and e) $R=$350~nm, all with $L=$5~nm. Dots show the full simulation results, while the closed symbols correspond to the real part of the semi-analytical LLD solutions. The lines are guides to the eye.
    }
    \label{fig:f_vs_D_RsL5}
\end{figure}

Fig.~\ref{fig:f_vs_D_RsL5} shows the radial spin wave eigenfrequencies and amplitudes versus~$D$ for a variety of disk sizes. In all cases, the frequency decreases as a function of increasing $D$, and the change is more dramatic for the higher order/shorter wavelength modes, as expected. The amplitude of the first mode decreases with increasing $D$. For the third mode, the amplitude drops slightly and then increases with increasing $D$, whereas the mode amplitudes increase for the even modes, especially the second mode. The amplitude changes are more pronounced for the larger disks. The trends observed in the simulations are also captured by the LLD results and the values are close, though the LLD method often leads to slightly higher frequencies.

\bibliographystyle{aipnum4-1}
\bibliography{references}

\begin{thebibliography}{24}%
\makeatletter
\providecommand \@ifxundefined [1]{%
 \@ifx{#1\undefined}
}%
\providecommand \@ifnum [1]{%
 \ifnum #1\expandafter \@firstoftwo
 \else \expandafter \@secondoftwo
 \fi
}%
\providecommand \@ifx [1]{%
 \ifx #1\expandafter \@firstoftwo
 \else \expandafter \@secondoftwo
 \fi
}%
\providecommand \natexlab [1]{#1}%
\providecommand \enquote  [1]{``#1''}%
\providecommand \bibnamefont  [1]{#1}%
\providecommand \bibfnamefont [1]{#1}%
\providecommand \citenamefont [1]{#1}%
\providecommand \href@noop [0]{\@secondoftwo}%
\providecommand \href [0]{\begingroup \@sanitize@url \@href}%
\providecommand \@href[1]{\@@startlink{#1}\@@href}%
\providecommand \@@href[1]{\endgroup#1\@@endlink}%
\providecommand \@sanitize@url [0]{\catcode `\\12\catcode `\$12\catcode
  `\&12\catcode `\#12\catcode `\^12\catcode `\_12\catcode `\%12\relax}%
\providecommand \@@startlink[1]{}%
\providecommand \@@endlink[0]{}%
\providecommand \url  [0]{\begingroup\@sanitize@url \@url }%
\providecommand \@url [1]{\endgroup\@href {#1}{\urlprefix }}%
\providecommand \urlprefix  [0]{URL }%
\providecommand \Eprint [0]{\href }%
\providecommand \doibase [0]{http://dx.doi.org/}%
\providecommand \selectlanguage [0]{\@gobble}%
\providecommand \bibinfo  [0]{\@secondoftwo}%
\providecommand \bibfield  [0]{\@secondoftwo}%
\providecommand \translation [1]{[#1]}%
\providecommand \BibitemOpen [0]{}%
\providecommand \bibitemStop [0]{}%
\providecommand \bibitemNoStop [0]{.\EOS\space}%
\providecommand \EOS [0]{\spacefactor3000\relax}%
\providecommand \BibitemShut  [1]{\csname bibitem#1\endcsname}%
\let\auto@bib@innerbib\@empty
\bibitem [{\citenamefont {Fert}, \citenamefont {Cros},\ and\ \citenamefont
  {Sampaio}(2013)}]{Fert2013SkyrmionsTrack}%
  \BibitemOpen
  \bibfield  {author} {\bibinfo {author} {\bibfnamefont {A.}~\bibnamefont
  {Fert}}, \bibinfo {author} {\bibfnamefont {V.}~\bibnamefont {Cros}}, \ and\
  \bibinfo {author} {\bibfnamefont {J.}~\bibnamefont {Sampaio}},\ }\href
  {\doibase 10.1038/nnano.2013.29} {\bibfield  {journal} {\bibinfo  {journal}
  {Nature Nanotechnology}\ }\textbf {\bibinfo {volume} {8}},\ \bibinfo {pages}
  {152} (\bibinfo {year} {2013})}\BibitemShut {NoStop}%
\bibitem [{\citenamefont {Rohart}\ and\ \citenamefont
  {Thiaville}(2013)}]{Rohart2013SkyrmionInteraction}%
  \BibitemOpen
  \bibfield  {author} {\bibinfo {author} {\bibfnamefont {S.}~\bibnamefont
  {Rohart}}\ and\ \bibinfo {author} {\bibfnamefont {A.}~\bibnamefont
  {Thiaville}},\ }\href {\doibase 10.1103/PhysRevB.88.184422} {\bibfield
  {journal} {\bibinfo  {journal} {Physical Review B}\ }\textbf {\bibinfo
  {volume} {88}},\ \bibinfo {pages} {184422} (\bibinfo {year}
  {2013})}\BibitemShut {NoStop}%
\bibitem [{\citenamefont {Woo}\ \emph {et~al.}(2016)\citenamefont {Woo},
  \citenamefont {Litzius}, \citenamefont {Kr{\"{u}}ger}, \citenamefont {Im},
  \citenamefont {Caretta}, \citenamefont {Richter}, \citenamefont {Mann},
  \citenamefont {Krone}, \citenamefont {Reeve}, \citenamefont {Weigand},
  \citenamefont {Agrawal}, \citenamefont {Lemesh}, \citenamefont {Mawass},
  \citenamefont {Fischer}, \citenamefont {Kl{\"{a}}ui},\ and\ \citenamefont
  {Beach}}]{Woo2016ObservationFerromagnets}%
  \BibitemOpen
  \bibfield  {author} {\bibinfo {author} {\bibfnamefont {S.}~\bibnamefont
  {Woo}}, \bibinfo {author} {\bibfnamefont {K.}~\bibnamefont {Litzius}},
  \bibinfo {author} {\bibfnamefont {B.}~\bibnamefont {Kr{\"{u}}ger}}, \bibinfo
  {author} {\bibfnamefont {M.~Y.}\ \bibnamefont {Im}}, \bibinfo {author}
  {\bibfnamefont {L.}~\bibnamefont {Caretta}}, \bibinfo {author} {\bibfnamefont
  {K.}~\bibnamefont {Richter}}, \bibinfo {author} {\bibfnamefont
  {M.}~\bibnamefont {Mann}}, \bibinfo {author} {\bibfnamefont {A.}~\bibnamefont
  {Krone}}, \bibinfo {author} {\bibfnamefont {R.~M.}\ \bibnamefont {Reeve}},
  \bibinfo {author} {\bibfnamefont {M.}~\bibnamefont {Weigand}}, \bibinfo
  {author} {\bibfnamefont {P.}~\bibnamefont {Agrawal}}, \bibinfo {author}
  {\bibfnamefont {I.}~\bibnamefont {Lemesh}}, \bibinfo {author} {\bibfnamefont
  {M.~A.}\ \bibnamefont {Mawass}}, \bibinfo {author} {\bibfnamefont
  {P.}~\bibnamefont {Fischer}}, \bibinfo {author} {\bibfnamefont
  {M.}~\bibnamefont {Kl{\"{a}}ui}}, \ and\ \bibinfo {author} {\bibfnamefont
  {G.~S.}\ \bibnamefont {Beach}},\ }\href {\doibase 10.1038/nmat4593}
  {\bibfield  {journal} {\bibinfo  {journal} {Nature Materials}\ }\textbf
  {\bibinfo {volume} {15}},\ \bibinfo {pages} {501} (\bibinfo {year}
  {2016})}\BibitemShut {NoStop}%
\bibitem [{\citenamefont {Moreau-Luchaire}\ \emph {et~al.}(2016)\citenamefont
  {Moreau-Luchaire}, \citenamefont {Moutafis}, \citenamefont {Reyren},
  \citenamefont {Sampaio}, \citenamefont {Vaz}, \citenamefont {Van~Horne},
  \citenamefont {Bouzehouane}, \citenamefont {Garcia}, \citenamefont
  {Deranlot}, \citenamefont {Warnicke}, \citenamefont {Wohlh{\"{u}}ter},
  \citenamefont {George}, \citenamefont {Weigand}, \citenamefont {Raabe},
  \citenamefont {Cros},\ and\ \citenamefont
  {Fert}}]{Moreau-Luchaire2016AdditiveTemperature}%
  \BibitemOpen
  \bibfield  {author} {\bibinfo {author} {\bibfnamefont {C.}~\bibnamefont
  {Moreau-Luchaire}}, \bibinfo {author} {\bibfnamefont {C.}~\bibnamefont
  {Moutafis}}, \bibinfo {author} {\bibfnamefont {N.}~\bibnamefont {Reyren}},
  \bibinfo {author} {\bibfnamefont {J.}~\bibnamefont {Sampaio}}, \bibinfo
  {author} {\bibfnamefont {C.~A.}\ \bibnamefont {Vaz}}, \bibinfo {author}
  {\bibfnamefont {N.}~\bibnamefont {Van~Horne}}, \bibinfo {author}
  {\bibfnamefont {K.}~\bibnamefont {Bouzehouane}}, \bibinfo {author}
  {\bibfnamefont {K.}~\bibnamefont {Garcia}}, \bibinfo {author} {\bibfnamefont
  {C.}~\bibnamefont {Deranlot}}, \bibinfo {author} {\bibfnamefont
  {P.}~\bibnamefont {Warnicke}}, \bibinfo {author} {\bibfnamefont
  {P.}~\bibnamefont {Wohlh{\"{u}}ter}}, \bibinfo {author} {\bibfnamefont
  {J.~M.}\ \bibnamefont {George}}, \bibinfo {author} {\bibfnamefont
  {M.}~\bibnamefont {Weigand}}, \bibinfo {author} {\bibfnamefont
  {J.}~\bibnamefont {Raabe}}, \bibinfo {author} {\bibfnamefont
  {V.}~\bibnamefont {Cros}}, \ and\ \bibinfo {author} {\bibfnamefont
  {A.}~\bibnamefont {Fert}},\ }\href {\doibase 10.1038/nnano.2015.313}
  {\bibfield  {journal} {\bibinfo  {journal} {Nature Nanotechnology}\ }\textbf
  {\bibinfo {volume} {11}},\ \bibinfo {pages} {444} (\bibinfo {year}
  {2016})}\BibitemShut {NoStop}%
\bibitem [{\citenamefont {Ryu}\ \emph {et~al.}(2013)\citenamefont {Ryu},
  \citenamefont {Thomas}, \citenamefont {Yang},\ and\ \citenamefont
  {Parkin}}]{Ryu2013ChiralWalls}%
  \BibitemOpen
  \bibfield  {author} {\bibinfo {author} {\bibfnamefont {K.-S.}\ \bibnamefont
  {Ryu}}, \bibinfo {author} {\bibfnamefont {L.}~\bibnamefont {Thomas}},
  \bibinfo {author} {\bibfnamefont {S.-H.}\ \bibnamefont {Yang}}, \ and\
  \bibinfo {author} {\bibfnamefont {S.}~\bibnamefont {Parkin}},\ }\href
  {\doibase 10.1038/nnano.2013.102} {\bibfield  {journal} {\bibinfo  {journal}
  {Nature Nanotechnology}\ }\textbf {\bibinfo {volume} {8}},\ \bibinfo {pages}
  {527} (\bibinfo {year} {2013})}\BibitemShut {NoStop}%
\bibitem [{\citenamefont {Emori}\ \emph {et~al.}(2013)\citenamefont {Emori},
  \citenamefont {Bauer}, \citenamefont {Ahn}, \citenamefont {Martinez},\ and\
  \citenamefont {Beach}}]{Emori2013Current-drivenWalls}%
  \BibitemOpen
  \bibfield  {author} {\bibinfo {author} {\bibfnamefont {S.}~\bibnamefont
  {Emori}}, \bibinfo {author} {\bibfnamefont {U.}~\bibnamefont {Bauer}},
  \bibinfo {author} {\bibfnamefont {S.-M.}\ \bibnamefont {Ahn}}, \bibinfo
  {author} {\bibfnamefont {E.}~\bibnamefont {Martinez}}, \ and\ \bibinfo
  {author} {\bibfnamefont {G.~S.~D.}\ \bibnamefont {Beach}},\ }\href {\doibase
  10.1038/nmat3675} {\bibfield  {journal} {\bibinfo  {journal} {Nature
  Materials}\ }\textbf {\bibinfo {volume} {12}},\ \bibinfo {pages} {611}
  (\bibinfo {year} {2013})}\BibitemShut {NoStop}%
\bibitem [{\citenamefont {DeJong}\ and\ \citenamefont
  {Livesey}(2015)}]{DeJong2015AnalyticAnisotropy}%
  \BibitemOpen
  \bibfield  {author} {\bibinfo {author} {\bibfnamefont {M.~D.}\ \bibnamefont
  {DeJong}}\ and\ \bibinfo {author} {\bibfnamefont {K.~L.}\ \bibnamefont
  {Livesey}},\ }\href {\doibase 10.1103/PhysRevB.92.214420} {\bibfield
  {journal} {\bibinfo  {journal} {Physical Review B}\ }\textbf {\bibinfo
  {volume} {92}},\ \bibinfo {pages} {214420} (\bibinfo {year}
  {2015})}\BibitemShut {NoStop}%
\bibitem [{\citenamefont {Moon}\ \emph {et~al.}(2013)\citenamefont {Moon},
  \citenamefont {Seo}, \citenamefont {Lee}, \citenamefont {Kim}, \citenamefont
  {Ryu}, \citenamefont {Lee}, \citenamefont {McMichael},\ and\ \citenamefont
  {Stiles}}]{Moon2013Spin-waveInteraction}%
  \BibitemOpen
  \bibfield  {author} {\bibinfo {author} {\bibfnamefont {J.~H.}\ \bibnamefont
  {Moon}}, \bibinfo {author} {\bibfnamefont {S.~M.}\ \bibnamefont {Seo}},
  \bibinfo {author} {\bibfnamefont {K.~J.}\ \bibnamefont {Lee}}, \bibinfo
  {author} {\bibfnamefont {K.~W.}\ \bibnamefont {Kim}}, \bibinfo {author}
  {\bibfnamefont {J.}~\bibnamefont {Ryu}}, \bibinfo {author} {\bibfnamefont
  {H.~W.}\ \bibnamefont {Lee}}, \bibinfo {author} {\bibfnamefont {R.~D.}\
  \bibnamefont {McMichael}}, \ and\ \bibinfo {author} {\bibfnamefont {M.~D.}\
  \bibnamefont {Stiles}},\ }\href {\doibase 10.1103/PhysRevB.88.184404}
  {\bibfield  {journal} {\bibinfo  {journal} {Physical Review B}\ }\textbf
  {\bibinfo {volume} {88}},\ \bibinfo {pages} {184404} (\bibinfo {year}
  {2013})}\BibitemShut {NoStop}%
\bibitem [{\citenamefont {Kostylev}(2014)}]{Kostylev2014InterfaceInteraction}%
  \BibitemOpen
  \bibfield  {author} {\bibinfo {author} {\bibfnamefont {M.}~\bibnamefont
  {Kostylev}},\ }\href {\doibase 10.1063/1.4883181} {\bibfield  {journal}
  {\bibinfo  {journal} {Journal of Applied Physics}\ }\textbf {\bibinfo
  {volume} {115}},\ \bibinfo {pages} {233902} (\bibinfo {year}
  {2014})}\BibitemShut {NoStop}%
\bibitem [{\citenamefont {Stashkevich}\ \emph {et~al.}(2015)\citenamefont
  {Stashkevich}, \citenamefont {Belmeguenai}, \citenamefont {Roussign{\'{e}}},
  \citenamefont {Cherif}, \citenamefont {Kostylev}, \citenamefont {Gabor},
  \citenamefont {Lacour}, \citenamefont {Tiusan},\ and\ \citenamefont
  {Hehn}}]{Stashkevich2015ExperimentalInteraction}%
  \BibitemOpen
  \bibfield  {author} {\bibinfo {author} {\bibfnamefont {A.~A.}\ \bibnamefont
  {Stashkevich}}, \bibinfo {author} {\bibfnamefont {M.}~\bibnamefont
  {Belmeguenai}}, \bibinfo {author} {\bibfnamefont {Y.}~\bibnamefont
  {Roussign{\'{e}}}}, \bibinfo {author} {\bibfnamefont {S.~M.}\ \bibnamefont
  {Cherif}}, \bibinfo {author} {\bibfnamefont {M.}~\bibnamefont {Kostylev}},
  \bibinfo {author} {\bibfnamefont {M.}~\bibnamefont {Gabor}}, \bibinfo
  {author} {\bibfnamefont {D.}~\bibnamefont {Lacour}}, \bibinfo {author}
  {\bibfnamefont {C.}~\bibnamefont {Tiusan}}, \ and\ \bibinfo {author}
  {\bibfnamefont {M.}~\bibnamefont {Hehn}},\ }\href {\doibase
  10.1103/PhysRevB.91.214409} {\bibfield  {journal} {\bibinfo  {journal}
  {Physical Review B}\ }\textbf {\bibinfo {volume} {91}},\ \bibinfo {pages}
  {214409} (\bibinfo {year} {2015})}\BibitemShut {NoStop}%
\bibitem [{\citenamefont {Nembach}\ \emph {et~al.}(2015)\citenamefont
  {Nembach}, \citenamefont {Shaw}, \citenamefont {Weiler}, \citenamefont
  {Ju{\'{e}}},\ and\ \citenamefont {Silva}}]{Nembach2015LinearFilms}%
  \BibitemOpen
  \bibfield  {author} {\bibinfo {author} {\bibfnamefont {H.~T.}\ \bibnamefont
  {Nembach}}, \bibinfo {author} {\bibfnamefont {J.~M.}\ \bibnamefont {Shaw}},
  \bibinfo {author} {\bibfnamefont {M.}~\bibnamefont {Weiler}}, \bibinfo
  {author} {\bibfnamefont {E.}~\bibnamefont {Ju{\'{e}}}}, \ and\ \bibinfo
  {author} {\bibfnamefont {T.~J.}\ \bibnamefont {Silva}},\ }\href {\doibase
  10.1038/nphys3418} {\bibfield  {journal} {\bibinfo  {journal} {Nature
  Physics}\ }\textbf {\bibinfo {volume} {11}},\ \bibinfo {pages} {825}
  (\bibinfo {year} {2015})}\BibitemShut {NoStop}%
\bibitem [{\citenamefont {Di}\ \emph {et~al.}(2015)\citenamefont {Di},
  \citenamefont {Zhang}, \citenamefont {Lim}, \citenamefont {Ng}, \citenamefont
  {Kuok}, \citenamefont {Yu}, \citenamefont {Yoon}, \citenamefont {Qiu},\ and\
  \citenamefont {Yang}}]{Di2015DirectFilm}%
  \BibitemOpen
  \bibfield  {author} {\bibinfo {author} {\bibfnamefont {K.}~\bibnamefont
  {Di}}, \bibinfo {author} {\bibfnamefont {V.~L.}\ \bibnamefont {Zhang}},
  \bibinfo {author} {\bibfnamefont {H.~S.}\ \bibnamefont {Lim}}, \bibinfo
  {author} {\bibfnamefont {S.~C.}\ \bibnamefont {Ng}}, \bibinfo {author}
  {\bibfnamefont {M.~H.}\ \bibnamefont {Kuok}}, \bibinfo {author}
  {\bibfnamefont {J.}~\bibnamefont {Yu}}, \bibinfo {author} {\bibfnamefont
  {J.}~\bibnamefont {Yoon}}, \bibinfo {author} {\bibfnamefont {X.}~\bibnamefont
  {Qiu}}, \ and\ \bibinfo {author} {\bibfnamefont {H.}~\bibnamefont {Yang}},\
  }\href {\doibase 10.1103/PhysRevLett.114.047201} {\bibfield  {journal}
  {\bibinfo  {journal} {Physical Review Letters}\ }\textbf {\bibinfo {volume}
  {114}},\ \bibinfo {pages} {047201} (\bibinfo {year} {2015})}\BibitemShut
  {NoStop}%
\bibitem [{\citenamefont {Ma}\ \emph {et~al.}(2017)\citenamefont {Ma},
  \citenamefont {Yu}, \citenamefont {Razavi}, \citenamefont {Sasaki},
  \citenamefont {Li}, \citenamefont {Hao}, \citenamefont {Tolbert},
  \citenamefont {Wang},\ and\ \citenamefont
  {Li}}]{Ma2017Dzyaloshinskii-MoriyaInterface}%
  \BibitemOpen
  \bibfield  {author} {\bibinfo {author} {\bibfnamefont {X.}~\bibnamefont
  {Ma}}, \bibinfo {author} {\bibfnamefont {G.}~\bibnamefont {Yu}}, \bibinfo
  {author} {\bibfnamefont {S.~A.}\ \bibnamefont {Razavi}}, \bibinfo {author}
  {\bibfnamefont {S.~S.}\ \bibnamefont {Sasaki}}, \bibinfo {author}
  {\bibfnamefont {X.}~\bibnamefont {Li}}, \bibinfo {author} {\bibfnamefont
  {K.}~\bibnamefont {Hao}}, \bibinfo {author} {\bibfnamefont {S.~H.}\
  \bibnamefont {Tolbert}}, \bibinfo {author} {\bibfnamefont {K.~L.}\
  \bibnamefont {Wang}}, \ and\ \bibinfo {author} {\bibfnamefont
  {X.}~\bibnamefont {Li}},\ }\href {\doibase 10.1103/PhysRevLett.119.027202}
  {\bibfield  {journal} {\bibinfo  {journal} {Physical Review Letters}\
  }\textbf {\bibinfo {volume} {119}},\ \bibinfo {pages} {027202} (\bibinfo
  {year} {2017})}\BibitemShut {NoStop}%
\bibitem [{\citenamefont {Zingsem}\ \emph {et~al.}(2019)\citenamefont
  {Zingsem}, \citenamefont {Farle}, \citenamefont {Stamps},\ and\ \citenamefont
  {Camley}}]{Zingsem2019UnusualWaves}%
  \BibitemOpen
  \bibfield  {author} {\bibinfo {author} {\bibfnamefont {B.~W.}\ \bibnamefont
  {Zingsem}}, \bibinfo {author} {\bibfnamefont {M.}~\bibnamefont {Farle}},
  \bibinfo {author} {\bibfnamefont {R.~L.}\ \bibnamefont {Stamps}}, \ and\
  \bibinfo {author} {\bibfnamefont {R.~E.}\ \bibnamefont {Camley}},\ }\href
  {\doibase 10.1103/PhysRevB.99.214429} {\bibfield  {journal} {\bibinfo
  {journal} {Physical Review B}\ }\textbf {\bibinfo {volume} {99}},\ \bibinfo
  {pages} {214429} (\bibinfo {year} {2019})}\BibitemShut {NoStop}%
\bibitem [{\citenamefont {Garcia-Sanchez}\ \emph {et~al.}(2014)\citenamefont
  {Garcia-Sanchez}, \citenamefont {Borys}, \citenamefont {Vansteenkiste},
  \citenamefont {Kim},\ and\ \citenamefont
  {Stamps}}]{Garcia-Sanchez2014NonreciprocalInteraction}%
  \BibitemOpen
  \bibfield  {author} {\bibinfo {author} {\bibfnamefont {F.}~\bibnamefont
  {Garcia-Sanchez}}, \bibinfo {author} {\bibfnamefont {P.}~\bibnamefont
  {Borys}}, \bibinfo {author} {\bibfnamefont {A.}~\bibnamefont
  {Vansteenkiste}}, \bibinfo {author} {\bibfnamefont {J.-V.}\ \bibnamefont
  {Kim}}, \ and\ \bibinfo {author} {\bibfnamefont {R.~L.}\ \bibnamefont
  {Stamps}},\ }\href {\doibase 10.1103/PhysRevB.89.224408} {\bibfield
  {journal} {\bibinfo  {journal} {Physical Review B}\ }\textbf {\bibinfo
  {volume} {89}},\ \bibinfo {pages} {224408} (\bibinfo {year}
  {2014})}\BibitemShut {NoStop}%
\bibitem [{\citenamefont {Mruczkiewicz}, \citenamefont {Krawczyk},\ and\
  \citenamefont {Guslienko}(2017)}]{Mruczkiewicz2017SpinStates}%
  \BibitemOpen
  \bibfield  {author} {\bibinfo {author} {\bibfnamefont {M.}~\bibnamefont
  {Mruczkiewicz}}, \bibinfo {author} {\bibfnamefont {M.}~\bibnamefont
  {Krawczyk}}, \ and\ \bibinfo {author} {\bibfnamefont {K.~Y.}\ \bibnamefont
  {Guslienko}},\ }\href {\doibase 10.1103/PhysRevB.95.094414} {\bibfield
  {journal} {\bibinfo  {journal} {Physical Review B}\ }\textbf {\bibinfo
  {volume} {95}},\ \bibinfo {pages} {094414} (\bibinfo {year}
  {2017})}\BibitemShut {NoStop}%
\bibitem [{\citenamefont {Novosad}\ \emph {et~al.}(2005)\citenamefont
  {Novosad}, \citenamefont {Fradin}, \citenamefont {Roy}, \citenamefont
  {Buchanan}, \citenamefont {Guslienko},\ and\ \citenamefont
  {Bader}}]{Novosad2005MagneticDots}%
  \BibitemOpen
  \bibfield  {author} {\bibinfo {author} {\bibfnamefont {V.}~\bibnamefont
  {Novosad}}, \bibinfo {author} {\bibfnamefont {F.~Y.}\ \bibnamefont {Fradin}},
  \bibinfo {author} {\bibfnamefont {P.~E.}\ \bibnamefont {Roy}}, \bibinfo
  {author} {\bibfnamefont {K.~S.}\ \bibnamefont {Buchanan}}, \bibinfo {author}
  {\bibfnamefont {K.~Y.}\ \bibnamefont {Guslienko}}, \ and\ \bibinfo {author}
  {\bibfnamefont {S.~D.}\ \bibnamefont {Bader}},\ }\href {\doibase
  10.1103/PhysRevB.72.024455} {\bibfield  {journal} {\bibinfo  {journal}
  {Physical Review B}\ }\textbf {\bibinfo {volume} {72}},\ \bibinfo {pages}
  {024455} (\bibinfo {year} {2005})}\BibitemShut {NoStop}%
\bibitem [{\citenamefont {Demokritov}\ and\ \citenamefont
  {Demidov}(2008)}]{Demokritov2008Micro-BrillouinNanostructures}%
  \BibitemOpen
  \bibfield  {author} {\bibinfo {author} {\bibfnamefont {S.}~\bibnamefont
  {Demokritov}}\ and\ \bibinfo {author} {\bibfnamefont {V.}~\bibnamefont
  {Demidov}},\ }\href {\doibase 10.1109/TMAG.2007.910227} {\bibfield  {journal}
  {\bibinfo  {journal} {IEEE Transactions on Magnetics}\ }\textbf {\bibinfo
  {volume} {44}},\ \bibinfo {pages} {6} (\bibinfo {year} {2008})}\BibitemShut
  {NoStop}%
\bibitem [{\citenamefont {Vogt}\ \emph {et~al.}(2011)\citenamefont {Vogt},
  \citenamefont {Sukhostavets}, \citenamefont {Schultheiss}, \citenamefont
  {Obry}, \citenamefont {Pirro}, \citenamefont {Serga}, \citenamefont
  {Sebastian}, \citenamefont {Gonzalez}, \citenamefont {Guslienko},\ and\
  \citenamefont {Hillebrands}}]{Vogt2011OpticalDisks}%
  \BibitemOpen
  \bibfield  {author} {\bibinfo {author} {\bibfnamefont {K.}~\bibnamefont
  {Vogt}}, \bibinfo {author} {\bibfnamefont {O.}~\bibnamefont {Sukhostavets}},
  \bibinfo {author} {\bibfnamefont {H.}~\bibnamefont {Schultheiss}}, \bibinfo
  {author} {\bibfnamefont {B.}~\bibnamefont {Obry}}, \bibinfo {author}
  {\bibfnamefont {P.}~\bibnamefont {Pirro}}, \bibinfo {author} {\bibfnamefont
  {A.~A.}\ \bibnamefont {Serga}}, \bibinfo {author} {\bibfnamefont
  {T.}~\bibnamefont {Sebastian}}, \bibinfo {author} {\bibfnamefont
  {J.}~\bibnamefont {Gonzalez}}, \bibinfo {author} {\bibfnamefont {K.~Y.}\
  \bibnamefont {Guslienko}}, \ and\ \bibinfo {author} {\bibfnamefont
  {B.}~\bibnamefont {Hillebrands}},\ }\href {\doibase
  10.1103/PhysRevB.84.174401} {\bibfield  {journal} {\bibinfo  {journal}
  {Physical Review B}\ }\textbf {\bibinfo {volume} {84}},\ \bibinfo {pages}
  {174401} (\bibinfo {year} {2011})}\BibitemShut {NoStop}%
\bibitem [{\citenamefont {Giovannini}\ \emph {et~al.}(2004)\citenamefont
  {Giovannini}, \citenamefont {Montoncello}, \citenamefont {Nizzoli},
  \citenamefont {Gubbiotti}, \citenamefont {Carlotti}, \citenamefont {Okuno},
  \citenamefont {Shinjo},\ and\ \citenamefont
  {Grimsditch}}]{Giovannini2004SpinStates}%
  \BibitemOpen
  \bibfield  {author} {\bibinfo {author} {\bibfnamefont {L.}~\bibnamefont
  {Giovannini}}, \bibinfo {author} {\bibfnamefont {F.}~\bibnamefont
  {Montoncello}}, \bibinfo {author} {\bibfnamefont {F.}~\bibnamefont
  {Nizzoli}}, \bibinfo {author} {\bibfnamefont {G.}~\bibnamefont {Gubbiotti}},
  \bibinfo {author} {\bibfnamefont {G.}~\bibnamefont {Carlotti}}, \bibinfo
  {author} {\bibfnamefont {T.}~\bibnamefont {Okuno}}, \bibinfo {author}
  {\bibfnamefont {T.}~\bibnamefont {Shinjo}}, \ and\ \bibinfo {author}
  {\bibfnamefont {M.}~\bibnamefont {Grimsditch}},\ }\href {\doibase
  10.1103/PhysRevB.70.172404} {\bibfield  {journal} {\bibinfo  {journal}
  {Physical Review B}\ }\textbf {\bibinfo {volume} {70}},\ \bibinfo {pages}
  {172404} (\bibinfo {year} {2004})}\BibitemShut {NoStop}%
\bibitem [{\citenamefont {Guslienko}\ \emph {et~al.}(2005)\citenamefont
  {Guslienko}, \citenamefont {Scholz}, \citenamefont {Chantrell},\ and\
  \citenamefont {Novosad}}]{Guslienko2005Vortex-stateDots}%
  \BibitemOpen
  \bibfield  {author} {\bibinfo {author} {\bibfnamefont {K.~Y.}\ \bibnamefont
  {Guslienko}}, \bibinfo {author} {\bibfnamefont {W.}~\bibnamefont {Scholz}},
  \bibinfo {author} {\bibfnamefont {R.~W.}\ \bibnamefont {Chantrell}}, \ and\
  \bibinfo {author} {\bibfnamefont {V.}~\bibnamefont {Novosad}},\ }\href
  {\doibase 10.1103/PhysRevB.71.144407} {\bibfield  {journal} {\bibinfo
  {journal} {Physical Review B}\ }\textbf {\bibinfo {volume} {71}},\ \bibinfo
  {pages} {144407} (\bibinfo {year} {2005})}\BibitemShut {NoStop}%
\bibitem [{\citenamefont {Zaspel}\ \emph {et~al.}(2009)\citenamefont {Zaspel},
  \citenamefont {Wright}, \citenamefont {Galkin},\ and\ \citenamefont
  {Ivanov}}]{Zaspel2009FrequenciesDisks}%
  \BibitemOpen
  \bibfield  {author} {\bibinfo {author} {\bibfnamefont {C.~E.}\ \bibnamefont
  {Zaspel}}, \bibinfo {author} {\bibfnamefont {E.~S.}\ \bibnamefont {Wright}},
  \bibinfo {author} {\bibfnamefont {A.~Y.}\ \bibnamefont {Galkin}}, \ and\
  \bibinfo {author} {\bibfnamefont {B.~A.}\ \bibnamefont {Ivanov}},\ }\href
  {\doibase 10.1103/PhysRevB.80.094415} {\bibfield  {journal} {\bibinfo
  {journal} {Physical Review B}\ }\textbf {\bibinfo {volume} {80}},\ \bibinfo
  {pages} {094415} (\bibinfo {year} {2009})}\BibitemShut {NoStop}%
\bibitem [{\citenamefont {Fischbacher}\ \emph {et~al.}(2017)\citenamefont
  {Fischbacher}, \citenamefont {Kovacs}, \citenamefont {Oezelt}, \citenamefont
  {Schrefl}, \citenamefont {Exl}, \citenamefont {Fidler}, \citenamefont
  {Suess}, \citenamefont {Sakuma}, \citenamefont {Yano}, \citenamefont {Kato},
  \citenamefont {Shoji},\ and\ \citenamefont
  {Manabe}}]{Fischbacher2017NonlinearMicromagnetics}%
  \BibitemOpen
  \bibfield  {author} {\bibinfo {author} {\bibfnamefont {J.}~\bibnamefont
  {Fischbacher}}, \bibinfo {author} {\bibfnamefont {A.}~\bibnamefont {Kovacs}},
  \bibinfo {author} {\bibfnamefont {H.}~\bibnamefont {Oezelt}}, \bibinfo
  {author} {\bibfnamefont {T.}~\bibnamefont {Schrefl}}, \bibinfo {author}
  {\bibfnamefont {L.}~\bibnamefont {Exl}}, \bibinfo {author} {\bibfnamefont
  {J.}~\bibnamefont {Fidler}}, \bibinfo {author} {\bibfnamefont
  {D.}~\bibnamefont {Suess}}, \bibinfo {author} {\bibfnamefont
  {N.}~\bibnamefont {Sakuma}}, \bibinfo {author} {\bibfnamefont
  {M.}~\bibnamefont {Yano}}, \bibinfo {author} {\bibfnamefont {A.}~\bibnamefont
  {Kato}}, \bibinfo {author} {\bibfnamefont {T.}~\bibnamefont {Shoji}}, \ and\
  \bibinfo {author} {\bibfnamefont {A.}~\bibnamefont {Manabe}},\ }\href
  {\doibase 10.1063/1.4981902} {\bibfield  {journal} {\bibinfo  {journal} {AIP
  Advances}\ }\textbf {\bibinfo {volume} {7}},\ \bibinfo {pages} {045310}
  (\bibinfo {year} {2017})}\BibitemShut {NoStop}%
\bibitem [{\citenamefont {Vansteenkiste}\ \emph {et~al.}(2014)\citenamefont
  {Vansteenkiste}, \citenamefont {Leliaert}, \citenamefont {Dvornik},
  \citenamefont {Helsen}, \citenamefont {Garcia-Sanchez},\ and\ \citenamefont
  {Van~Waeyenberge}}]{Vansteenkiste2014TheMuMax3}%
  \BibitemOpen
  \bibfield  {author} {\bibinfo {author} {\bibfnamefont {A.}~\bibnamefont
  {Vansteenkiste}}, \bibinfo {author} {\bibfnamefont {J.}~\bibnamefont
  {Leliaert}}, \bibinfo {author} {\bibfnamefont {M.}~\bibnamefont {Dvornik}},
  \bibinfo {author} {\bibfnamefont {M.}~\bibnamefont {Helsen}}, \bibinfo
  {author} {\bibfnamefont {F.}~\bibnamefont {Garcia-Sanchez}}, \ and\ \bibinfo
  {author} {\bibfnamefont {B.}~\bibnamefont {Van~Waeyenberge}},\ }\href
  {\doibase 10.1063/1.4899186} {\bibfield  {journal} {\bibinfo  {journal} {AIP
  Advances}\ }\textbf {\bibinfo {volume} {4}},\ \bibinfo {pages} {107133}
  (\bibinfo {year} {2014})}\BibitemShut {NoStop}%
\end{thebibliography}%

%
%
%
\end{document}